\documentclass[12pt]{iopart}
\pdfoutput=1
\usepackage{iopams}  
\usepackage{graphicx}
\usepackage{setstack}
\usepackage{color}
\usepackage{footmisc}

\usepackage{braket}
\usepackage{bbm}

\newcommand{\eqref}[1]{(\ref{#1})}
\begin{document}\title{Dynamic Variational Study of Chaos: Spin Glasses in Three Dimensions}

\author{A.~Billoire$^{1}$, L.~A.~Fernandez$^{2,3}$, A.~Maiorano$^{4,3}$, E.~Marinari$^{4,5,6}$, V.~Martin-Mayor$^{2,3}$, J.~Moreno-Gordo$^{3,7}$, G.~Parisi$^{4,5,6}$, F.~Ricci-Tersenghi $^{4,5,6}$,
J.~J.~Ruiz-Lorenzo$^{8,3}$.}

\address{$^1$ Institut de Physique Th\'eorique,
  CEA Saclay and CNRS, 91191 Gif-sur-Yvette, France.}
\address{$^2$ Depto. F\'{\i}sica Te\'orica I. Facultad de Ciencias
  F\'{\i}sicas. Universidad Complutense de Madrid. Madrid 28040. Spain.}
\address{$^3$ Instituto de Biocomputaci\'on y
  F\'{\i}sica de Sistemas Complejos (BIFI), 50018 Zaragoza, Spain.}
\address{$^4$ Dipartimento di Fisica, Sapienza
  Universit\`a di Roma, I-00185 Rome, Italy.}
\address{$^5$ Nanotec, Consiglio Nazionale delle Ricerche, I-00185 Rome, Italy.}
\address{$^6$ Istituto Nazionale di Fisica Nucleare, Sezione di Roma 1, I-00185 Rome, Italy.}
\address{$^7$ Departamento de F\'{\i}sica Te\'orica, Universidad de Zaragoza, 50009 Zaragoza, Spain.}
\address{$^8$ Departamento de F\'{\i}sica and
  Instituto de Computaci\'on Cient\'{\i}fica Avanzada (ICCAEx), Universidad de
  Extremadura, 06071 Badajoz, Spain.}

\date{\today}

\begin{abstract}

We have introduced a variational method to improve the computation of
integrated correlation times in the Parallel Tempering Dynamics,
obtaining a better estimate (a lower bound, at least) of the
exponential correlation time. Using this determination of the
correlation times, we revisited the problem of the characterization of
the chaos in temperature in finite dimensional spin glasses spin by
means of the study of correlations between different chaos indicators
computed in the static and the correlation times of the Parallel
Tempering dynamics. The sample-distribution of the characteristic
time for the Parallel Tempering dynamics turns out to be fat-tailed
 and it obeys finite-size scaling.
\end{abstract}

\maketitle

\section{Introduction}

Markov-chain Monte Carlo methods (MC) are a crucial tool to study
non-perturbative problems in statistical mechanics and quantum field
theory~\cite{sokal:92,sokal:97,newman:99,landau:05}. A major problem
arises however when studying systems with a rugged free-energy
landscapes: we have in mind for example
spin-glasses~\cite{mezard:87,young:98}, or glass-forming
liquids~\cite{cavagna:09}). The presence of many free-energy local
minima often causes the numerical simulation to get trapped and, as a
consequence, does not allow a correct sampling of the phase space.

The parallel tempering method, connected to the original simulated
tempering method \cite{marinari:92} and also known as the replica
exchange method, was devised to overcome these
difficulties~\cite{geyer:91, hukushima:96, marinari:98b}.  One
considers $N_T$ copies (or clones) of the system, and uses for each of them a
different temperature $T_i$, with $T_1<T_2<\cdots<T_{N_T}$.  As
explained in~\ref{app:Markov}, the target probability
distribution for the $N_T$ systems is the product of the Boltzmann
distributions at the different temperatures.  A parallel tempering
numerical simulation is based on two alternating sets of steps. First,
each system copy undergoes independently a standard Monte Carlo
dynamics (for example Metropolis) at its own temperature: one can use
each time one or more MC steps. Second, pairs of spin configurations
attempt to exchange their temperatures.\footnote{The
  temperature-exchange rule is designed to have the target probability
  distribution as the unique equilibrium measure. In other words the
  restriction of the total measure to a single temperature is exactly
  the appropriate Boltzmann distribution at that temperature, see~\ref{app:Markov}.}

The rationale behind parallel tempering is simple. Each system copy
undergoes a random walk in temperature space. When a system copy is at
a low temperature, it only explores the nearby free-energy local
minima. Instead, when its temperature is high, free-energy barriers
disappear: the copy can freely wander in phase space, and when it will
cool again it will typically fall in a different free energy valley,
with different local minima.  For parallel tempering to effectively
thermalize, it is crucial that any copy of the system spends its time
roughly evenly at every temperature: high temperatures are needed for
visiting all the phase space, low temperatures are needed to visit its
low free energy regions.  In fact parallel tempering is at this point
used in a very large number of very different applications (for
example in physics, biology, chemistry, engineering, statistics), and
considerable efforts have been devoted to improve it, from different
communities. Different temperature-exchange rules have been developed
and tested~\cite{sugita:99, calvo:05,earl:05, brenner:07, bittner:08,
  malakis:13}. Furthermore, it has been suggested that a significant
gain can be achieved by optimizing the choice of the $N_T$
temperatures~\cite{sabo:08,katzgraber:06b}.

In order to assess the relative merits of the above suggestions, one
needs a quantitative method. The theory of Markov chains suggest
considering the \emph{exponential} autocorrelation time
$\tau_{\mathrm{exp}}$ of the Monte Carlo dynamics as a relevant figure
of merit~\cite{sokal:97}. $\tau_{\mathrm{exp}}$ tells us how long we
should wait before equilibrium is reached.  Unfortunately,
$\tau_{\mathrm{exp}}$ is an elusive quantity.  In the context of a
parallel tempering simulation, it has been suggested that
$\tau_{\mathrm{exp}}$ is best computed by studying the
temperature-flow of the system copies~\cite{fernandez:09b,janus:10}:
the exchange of temperatures is indeed the slow mode of the combined
numerical simulation based on parallel tempering and Metropolis moves,
and it is the interesting process to quantify.  We will focus here in
the determination of $\tau_{\mathrm{exp}}$ for a parallel tempering
simulation of a spin-glass. Our choice entails no generality loss,
because the problem of finding the ground state (or low-temperature
configurations) in a spin-glass is NP-complete~\cite{barahona:82b}:
understanding it shed light on a large class of very interesting
phenomena. Furthermore, spin glasses show very clearly the major
problems that a parallel tempering simulation faces.

To be specific, we shall be considering the three dimensional
Edwards-Anderson model~\cite{edwards:75,edwards:76}. Ising variables
($s_i=\pm 1$) occupy the nodes of a cubic lattice of size $L$ with
periodic boundary conditions. Spins interact with their nearest
lattice-neighbors through the Hamiltonian
\begin{equation}\label{eq:H}
H = - \sum_{\braket{i,j}} J_{ij}s_i s_j \, ,
\end{equation}
where the quenched couplings $J_{ij}$ are drawn from a
bimodal probability distribution (so that $J_{ij}=\pm 1$ with $1/2$
probability) at the beginning of the simulation. A choice of couplings
$\{J_{ij}\}$ will be called a (disorder) sample (or realization)
hereafter.

A major complication in the numerical study of the
Hamiltonian~\eqref{eq:H} is that a large number of samples of the
system (the larger, the better) needs to be studied due to non the
self averaging property of the system.\footnote{Strictly speaking,
  non-self-averaging occurs only when the correlation length becomes
  of the order of magnitude of the system size (which is usually the
  case at the temperatures of interest).} Besides, below the critical
temperature $T_\mathrm{c}$, the value of $\tau_{\mathrm{exp}}$ (i.e.
the computational difficulty that characterizes the physical system)
presents huge sample to sample
fluctuations~\cite{fernandez:09b,janus:10} (see also
Fig.~\ref{fig:selection_samples_dinamica}). The presence of these
fluctuations makes the problem of computing $\tau_{\mathrm{exp}}$ very
relevant for saving CPU time (allowing, in this way, larger and more
accurate simulations for a given cost): knowing the value of
$\tau_{\mathrm{exp}}$ for each individual sample allows to save huge
amount of computer time by setting the chain length for a given sample
proportional to its own $\tau_{\mathrm{exp}}$.

There is a fairly general physical mechanism behind the dramatic
dispersion of $\tau_{\mathrm{exp}}$ (and behind its severe growth with
the system size), the so called
temperature chaos \cite{mckay:82,bray:87b,banavar:87,kondor:89,kondor:93,billoire:00,rizzo:01,mulet:01,billoire:02,krzakala:02,rizzo:03,sasaki:05,katzgraber:07,parisi:10,fernandez:13,billoire:14,wang:15,fernandez:16}.
Temperature chaos consists of a major reorganization of the typical
equilibrium configurations upon tiny temperature changes. A detailed
inspection shows how the effect arises on finite
systems~\cite{fernandez:13,martin-mayor:15,fernandez:16}.  Indeed, for
some samples one encounters chaotic-events taking place at well
defined temperatures, in the form of major changes of the spin
configurations as the temperature is lowered. Chaotic events are reminiscent of first-order phase
transitions (rounded in a finite system). In a fixed temperature
interval, $T_A<T<T_B$ with $T_B<T_\mathrm{c}$, a given sample may
undergo zero, one or even more chaotic events (the temperature
location of the chaotic events is also random). Given $T_A<T<T_B$, the
larger the system the larger is the probability of finding samples
displaying chaotic events in that temperature
region~\cite{fernandez:13}. Lowering $T_A$, while
keeping the size fixed, increases as well the probability of
encountering a chaotic event.

As it is intuitively obvious, temperature chaos turns out to be a
major obstacle for the parallel tempering temperature
flow~\cite{janus:10,fernandez:13,martin-mayor:15,fernandez:16}.  The
main point is that equilibrium in parallel tempering implies
equilibrium~\emph{at all temperatures}. Now, let us assume that the
typical equilibrium spin-configurations at two neighboring
temperatures in the temperature grid are vastly different. Clearly, if
one spin configuration \emph{of the low-temperature type} is
momentarily placed at the high temperature, it will have a hard time
traveling to the highest temperatures in temperature grid (because the
clones at the higher temperatures are fitter and the local spin-flip
dynamics is obviously inefficient to remediate this problem).
Furthermore, temperature chaos is relevant in the analysis of crucial
experimental results \cite{jonason:98, bellon:00, vincent:00,
  bouchaud:01b,ozon:03, yardimci:03,mueller:04, guchhait:15b} and in
the performance analysis of commercial quantum
annealers~\cite{martin-mayor:15,katzgraber:15,marshall:16}.
 
Here we revisit the problem of estimating $\tau_{\mathrm{exp}}$ and we
present a variational method that can potentially save a large amount
of computation time. Very often a numerical simulation needs to be
extended just because of the difficulties encountered in the
computation of $\tau_{\mathrm{exp}}$. Having in our hands a safe
mechanisms to estimate $\tau_{\mathrm{exp}}$ in an automated way (the
number of samples needed in a state of the art numerical simulation goes
by the thousands) can avoid unnecessary extensions of the simulation
length. We also investigate further the relationship between
temperature chaos, which is a \emph{static} equilibrium feature, and
$\tau_{\mathrm{exp}}$ which characterizes a Markov chain
\emph{dynamics}.

This paper is organized as follows. In section~\ref{sect:tau-def} we
introduce two different time scales that characterize a Monte Carlo
Markov chain.  Our simulations are described in
Sect.~\ref{sect:simulations}. We present our characterization of
temperature chaos in Sect.~\ref{sect:T-chaos}. The variational method
for the computation of the autocorrelation time $\tau_{\mathrm{exp}}$
is discussed in Sect.~\ref{sect:tau-var}.
Section~\ref{sect:dynamic_scaling} is devoted to the study of the
scaling properties of the Parallel Tempering method
$\tau_{\mathrm{exp}}$. We study in a very precise detail the
thermodynamic equilibrium features that characterize temperature
chaos~\cite{fernandez:13} in Sect.~\ref{sect:static-chaos}.  The
relationship between static and dynamic chaos indicators is studied in
Sect.~\ref{sect:static-dynamic-chaos}. Our discussion of results
concludes the paper in Sect.~\ref{sect:conclusions}. We provide in
\ref{app:simulation_parameters} a description of the
parameters of our simulations. In \ref{app:selection_parameters} we discuss the particular choice of
samples we use~\cite{billoire:17}. In 
\ref{app:geometry_S_MSC} we describe in detail the geometry used in
our implementation of the synchronous multispin coding. In
\ref{app:quatities_not_chaos} we discuss some quantities
which are surprisingly unrelated to chaos. Finally, in~\ref{app:Markov} we discuss in some detail the relationship
between time-correlations and system equilibration.

\section{Time scales in a Markov chain}\label{sect:tau-def}

This section is  a quick reminder of some basic
concepts. The interested reader is referred to Ref.~\cite{sokal:97}
for further details. Specific examples and computational recipes will
be discussed in Sect.~\ref{sect:tau-var} (see also~\ref{app:Markov}).

Almost all the Monte Carlo methods used in Statistical Physics are
based on the theory of Markov chains. A Markov chain starts from some
initial configuration and we need to know how long the Markov dynamics
must be run in order to reach equilibrium. This time scale is the
exponential autocorrelation time ($\tau_{\mathrm{exp}}$). In addition to
this time scale, for any physical quantity $f$ we can define a second
time scale, the integrated autocorrelation time
($\tau_{\mathrm{int},f}$), which controls statistical errors in
measuring $f$: two already equilibrated configurations whose time
difference is $2\tau_{\mathrm{int},f}$ are statistically independent in
an effective sense (but only as far as the quantity $f$ is concerned).

Under very mild assumptions (see below) it is possible to show that
the following inequality holds for any $f$:
\begin{equation}
\label{eq:exp-int}
\tau_{\mathrm{int},f}\le
\tau_\mathrm{exp} \,.
\end{equation}
The crucial point is that $\tau_{\mathrm{int},f}$ is relatively easy to
compute. Instead $\tau_\mathrm{exp}$ is rather elusive. Hence, we shall
use Eq.~\eqref{eq:exp-int} for a variational method analogous to the
Rayleigh-Ritz variational principle in Quantum Mechanics. In
Sect.~\ref{sect:tau-var}, we shall try different quantities $f$ and
compute $\tau_{\mathrm{int},f}$ for each of them. The largest value 
of $\tau_{\mathrm{int},f}$ will be our variational estimate for
$\tau_\mathrm{exp}$.

Let us recall that the equilibrium autocorrelation function for quantity
$f$   is
\begin{equation}\label{eq:def-Cf}
C_f(t)=E\left[ f(t_1) f(t_2)\right] - E\left[f(t_1)\right]^2\,,\quad t=t_1-t_2\,,
\end{equation}
where $E[\ldots]$ stands for the expectation value and the two times
$t_1$ and $t_2$ are large enough to reach equilibrium [hence
  $E\left[f(t_1)\right]=E\left[f(t_2)\right]$ and $C_f(t)=C_f(-t)$].
The integrated autocorrelation time is defined from the normalized
correlation function $\hat C_f(t)$:
\begin{equation}\label{eq:def-tauint}
\hat C_f(t)\equiv \frac{C_f(t)}{C_f(0)}\,,\quad 
\tau_{\mathrm{int},f}=\frac{1}{2}+\sum_{t=1}^\infty \hat C_f(t)\,.
\end{equation}
The normalized autocorrelation function can be expressed in terms of
the eigenvalues $\lambda_n$ of the transition probability matrix
projected on the subspace orthogonal to its eigenvalue 1
($1>|\lambda_1|\geq |\lambda_2|\geq\ldots$), see Ref.~\cite{sokal:97},
\begin{equation}\label{eq:eigen-1}
\hat C_f(t)=\sum_n A_{n,f} \lambda_n^{|t|}\,,\quad \sum_n A_{n,f}=1\,,
\end{equation}
where the index $n$ runs from 1 to $N_T!2^{N_TL^D} - 1$, in our case.

The amplitudes $A_{n,f}$  depend on $f$, while the $\lambda_n$
are $f$-independent. In terms of the $A_{n,f}$'s and $\lambda_n$'s one has
\begin{equation}\label{eq:eigen-2}
\tau_{\mathrm{int},f} =\frac{1}{2}\ +\ \sum_n\, A_{n,f}\, \frac{\lambda_n}{1-\lambda_n}\,.
\end{equation}
Now, in practical applications the (leading) $A_{n,f}$'s and
$\lambda_n$'s are real positive. Hence,
$\lambda_n=\mathrm{e}^{-1/\tau_n}$ defines the characteristic time
$\tau_n$. The exponential autocorrelation time of the Markov chain
$\tau_\mathrm{exp}$ is just $\tau_1$, the largest of the $\tau_n$. Now,
for $\tau_n\gg 1$ one has $\lambda_n/(1-\lambda_n)=\tau_n+{\cal
  O}(1/\tau_n)$ and Eqs.~(\ref{eq:eigen-1},\ref{eq:eigen-2}) become
\begin{equation}\label{eq:eigen-3}
\hat C_f(t)=\sum_n A_{n,f} \mathrm{e}^{-|t|/\tau_n}\,,\quad \tau_{\mathrm{int},f} =\frac{1}{2}+\sum_n A_{n,f} \tau_n\,.
\end{equation}
The variational method in Eq.~\eqref{eq:exp-int} follows immediately
from Eq.~\eqref{eq:eigen-3}. The optimal choice for the observable $f$
would have $A_{1,f}=1$ (and $A_{n>1,f}=0$) in its decomposition in
characteristic times.

\section{Numerical Simulations}\label{sect:simulations}

We develop our study in the context of Ref.~\cite{billoire:17}, in
which the metastate was studied. For this reason, our realizations of
disorder $\{J_{ij}\}$ ({samples}) are particular. In 
\ref{app:simulation_parameters} and \ref{app:selection_parameters} we
explain how the samples have been chosen and argue that this choice
does not affect the results.

We have simulated this model using the Parallel Tempering method with
Metropolis updates. See \ref{app:simulation_parameters}
 and \ref{app:selection_parameters} for the reasons behind our choice 
of the minimal temperature in the  Parallel
Tempering. Regarding the Metropolis updates we have used either the
multisample multispin coding (MUSA-MSC)~\cite{newman:99} or the
multisite multispin coding (MUSI-MSC)~\cite{fernandez:15} techniques,
that we will briefly describe.

Intel and AMD CPUs support 128 and 256-bit words in their streaming
extensions.  It is known that we can perform the Metropolis update of
a single spin by using a sequence of Boolean operations
\cite{newman:99}, so we can take advantage of current CPU technology
to simulate 128 or 256 systems simultaneously. This method is widely
used in computational physics
\cite{manssen:15,janus:13,lulli:15b,leuzzi:08,banos:12,fernandez:09f,newman:99}
and it is denominated multisample multispin coding (MUSA-MSC). The
most efficient version of our MUSA-MSC code turned out to be the one
with 128 bits.

However, there exist certain samples with such a sluggish dynamic that
MUSA-MSC ceases to be efficient. Indeed, if only a few of the 128
samples coded in a computer word are not yet thermalized, continuing
the simulation of the already equilibrated samples is a waste of
computer time. This problem is particularly acute for $L=16$ and $24$,
because the width of the autocorrelation time distribution increases
with $L$ (see Sect.~\ref{sect:dynamic_scaling}). For those misbehaving
instances we turned to multisite multispin coding (MUSI-MSC): the 256
bits in a computer word now code 256 distinct spins of a single
replica of a single sample~\cite{fernandez:15}.  In this way, we execute
the Metropolis algorithm in $L^3/256$ steps. Our implementation for
$L=24$ use a geometric arrangement differing from
Ref.~\cite{fernandez:15}, as explained in~\ref{app:geometry_S_MSC}. 

The simulations were carried out using either Intel Xeon E5-2680 or
AMD Opteron Processor 6272 processors. 12800 samples were simulated
(and 4 replicas per sample). More details of the simulations are given
in~\ref{app:simulation_parameters}.

\section{Characterizations of Temperature Chaos}\label{sect:T-chaos}

Temperature chaos will be studied from two complementary
viewpoints. The perspective offered by the Parallel Tempering dynamics
is considered in Sect.~\ref{sect:tau-var}.  The finite-size scaling of
the Parallel Tempering dynamics is studied in
Sect.~\ref{sect:dynamic_scaling}. The static viewpoint is considered
in Sect.~\ref{sect:static-chaos}. Finally, in
Sect.~\ref{sect:static-dynamic-chaos}, we will study the correlation
between the Parallel Tempering dynamics and temperature chaos.

\subsection{Dynamics: The Variational Method}\label{sect:tau-var}

Our scope here is to use Eq.~\eqref{eq:exp-int} in a variational
method to estimate the exponential autocorrelation time.  Consider the
eigenmode expansion in Eq.~\eqref{eq:eigen-3}. The optimal choice for
the observable $f$ would have $A_{1,f}=1$ (and $A_{n>1,f}=0$) in its
decomposition in characteristic times.\footnote{The reader is probably
  used to apply this formalism to the evolution of a single spin
  configuration. Here we shall need to enlarge this viewpoint to a
  parallel tempering simulation that involves several spin chains and
  to a function $f$ that is related to the temperature of a given
  chain. More details can be found in~\ref{app:Markov}.} We shall use
our physical intuition to approach this ideal.

As explained in the Introduction, the temperature chaos effect
suggests to focus our attention on the temperature flow along the
parallel tempering
dynamics~\cite{janus:10,martin-mayor:15,fernandez:16}. Let us consider
one of the $N_T$ system copies in the Parallel Tempering dynamics.
%The system copy can be at one of the allowed temperatures $T_1, T_2,\ldots T_{N_T}$. 
We shall describe the temperature random-walk through the
index $i_t$ that indicates that, at time $t$, our system copy is at
temperature $T_{i_t}$. The equilibrium probability for $i_t$ is just
the uniform probability over the set $\{1,2,\ldots,N_T\}$. If we consider
an arbitrary function of $i_t$ its equilibrium expectation value will be
\begin{equation}
E(f) = \frac{1}{N_T} \sum_{i = 0}^{N_T} f(i) \,.
\end{equation}
We shall consider as well \emph{pairs} of system copies. These pairs will be
described by two integers indices $i_t\neq j_t$. The equilibrium value of
an arbitrary function of a pair of system copies is
\begin{equation}
E(f) = \frac{1}{N_T (N_T-1)}\sum_{i=0}^{N_T} \sum_{j \neq i}^{N_T} f(i,j) \, . \label{eq:cond_norm2}
\end{equation}

We  will optimize three parameters: the type of function $f$, the
temperature $T^*$ where $f$ is zero, and a Wilson-Kadanoff
renormalization  block length, $l_\mathrm{blo}$. 
We will describe these three parameters in the next paragraphs.

We consider variational test-functions $f$ belonging to eight
different classes, see Table \ref{tab:table_functions}.  One of these
classes contains the linear functions studied in
Ref. \cite{janus:10}.  All our test-function have a vanishing
expectation value $E(f)=0$. We also request $f(T^*)=0$ for some $T^*
\in \{T_1,T_2,\ldots T_{N_T}\}$.  The location of $T^*$ is our second
variational parameter.  Specifically, our linear  
test-functions are
\begin{eqnarray}
T>T^*\!:\ f_{T^*}(T)&=&a_+\,(T-T^*)\,,\\
T<T^*\!:\ f_{T^*}(T)&=&a_-\,(T-T^*)\,.
\end{eqnarray}
We require the two amplitudes $a_+$ and $a_-$ to be positive. Their
ratio is fixed by imposing $E(f_{T^*})=0$. Indeed, we need to
fix only the ratio $a_+/a_-$, because the overall scale of the test
function $f_{T^*}$ is irrelevant. 
Besides, we consider quadratic ($p=2$) and cubic ($p=3$) test-functions:
\begin{eqnarray}
T>T^*\!:\,f_{T^*}(T)&\!=\!&a_+\,(T-T^*)^p (2 T_{N_T}-T^*\!-T),\\
T<T^*\!:\,f_{T^*}(T)&\!=\!&a_-\,(T^*\!-T)^p (2 T_1-T^*\!-T)\,.
\end{eqnarray}
We choose again $a_+,a_->0$, and the ratio $a_+/a_-$ is fixed by
imposing $E(f_{T^*})=0$. Note that all our test-functions are
continuous at $T^*$ (the cubic $f_{T^*}$ are even differentiable at $T^*$).

Now, for each $f$ and $T^*$, we need to estimate the
autocorrelation function $C_{f,T^*}(t)$, recall Eq.~\eqref{eq:def-Cf}, and
the related integrated autocorrelation time~\eqref{eq:def-tauint}. Let
$\tilde{f}_{T^*} \equiv f_{T^*} - E(f_{T^*})$. $C_{f,T^*}(t)$ is estimated as:
\begin{equation}
C_{f,T^*}(t) = \frac{n_\mathrm{Met}}{N_s - t_0 -t}\sum_{t'=t_0}^{N_s - t} \tilde{f}_{T^*}(i_{t'})\tilde{f}_{T^*}(i_{t'+t}) \, . \label{eq:correlacion}
\end{equation}
Here, $N_S$ is the number of times we stored the state of the PT
indices $i_t$ in the hard drive. Note that
$t_0$ must be much greater than $\tau_\mathrm{int}$, in order to be safely  in
the equilibrium regime. The parameter $n_\mathrm{Met}$ is the
periodicity with which we record the time indices $i_t$ (in most of
this work, $n_\mathrm{Met}=25000$ Metropolis sweeps). Note that
$C_{f,T^*}(t)$ is independent of the system copy. Therefore, we can
average over the $N_T$ numerical estimations of $C_{f,T^*}(t)$ (as
well as over the four independent replicas), which greatly enhances
the statistics. The computation for functions $f$ depending on a pair
of system copies is  analogous.

Once we have computed $C_{f,T^*}(t)$, the normalized correlation
function is just $\hat C_{f,T^*}(t)$, and the integrated
autocorrelation time can be computed in the standard
way~\cite{sokal:97}
\begin{equation}
\tau_{\mathrm{int},f,T^*} \approx n_\mathrm{Met} \left[\,\frac{1}{2} + \sum_{t=0}^{W} \hat{C_f}(t)\,\right] \, ,
\label{eq:tau}
\end{equation}
where $W$ is a self-consistent window~\cite{sokal:97} that avoids the
divergence of the variance of $\tau_{\mathrm{int},f,T^*}$ (we impose 
$\tau_{\mathrm{int},f,T^*}<10 W$).

%At this point, one would just compute $\tau_{\mathrm{int},f,T^*}$ for
%each of the 8 types of $f$ and each $T^*$, choosing the maximum of
%$\tau_{\mathrm{int},f,T^*}$.  

We have found it advantageous
to consider a third variational parameter $l_\mathrm{blo}$ that we now
describe.
We build Wilson-Kadanoff blocks: the Monte Carlo
sequence $f_{T^*}(i_1), f_{T^*}(i_2),\ldots f_{T^*}(i_{N_s})$ is
divided into blocks of $l_\mathrm{blo}$ consecutive data (see
e.g. Ref.~\cite{amit:05}). We take the average of the $f_{T^*}(i_t)$
within a single block. This operation defines a new sequence of
$N_S/l_\mathrm{blo}$ renormalized times, over which the integrated
autocorrelation time can be estimated just as we did for the original
data $l_\mathrm{blo}=1$. The estimated autocorrelation time
should be rescaled by $l_\mathrm{blo}$ in order to recover the original
time units. The purpose of the blocking is to reduce
high-frequency fluctuations. 

There is a danger in the use of Wilson-Kadanoff blocks, though.
Formula \ref{eq:tau} was obtained assuming that $
\tau_{\mathrm{int},f,T^*} $ is much larger that the time step in the
right hand side.  In fact, $l_\mathrm{blo}$ can be made much greater than the
$\tau_\mathrm{exp}$ that we aim to estimate.  As a consequence, the
renormalized correlation function will vanish for times $t\neq
0$. This means than the integrated autocorrelation time will be $1/2$
(over the renormalized time-mesh). When turning back to physical time
units we shall find $\tau_\mathrm{int}=n_\mathrm{Met} \>l_\mathrm{blo}/2$
which diverges for large $l_\mathrm{blo}$. Hence, we need a practical
way to ensure that $l_\mathrm{blo}$ is not so large that all the
physical information has been erased. Our solution imposes
\begin{equation}
\tau_{\mathrm{int},f,T^*,l_\mathrm{blo}} < \frac{5}{2} n_\mathrm{Met} \> l_\mathrm{blo} \,,
\end{equation}
in order to consider the results of a given $l_\mathrm{blo}$.

We obtain for each sample, a huge number of values of
$\tau_\mathrm{int}$ corresponding to the eight different functions and the
different choices of $T^*$. We have tried for $T^*$  all the
temperatures $T_i$ in the lower half of the set of temperatures in our
Parallel Tempering simulation. The values of $l_\mathrm{blo}$ are taken
from the list $\{1,2,5,10,20,50,100,200,500,1000,2000\}$.

Our variational estimate $ \tau_\mathrm{int,var}$
is the largest of these numbers. This is a robust estimate (i.e. this
methodology does not provide spurious values) and therefore can be
implemented in an automatic way in the analysis, and allows for a
precise estimate of the needed thermalization time. 

We shall also consider below the temperature $T_d$ which is the $T^*$
for which the variational maximum is attained.

\begin{figure}
\centering
\includegraphics[width=0.8\columnwidth]{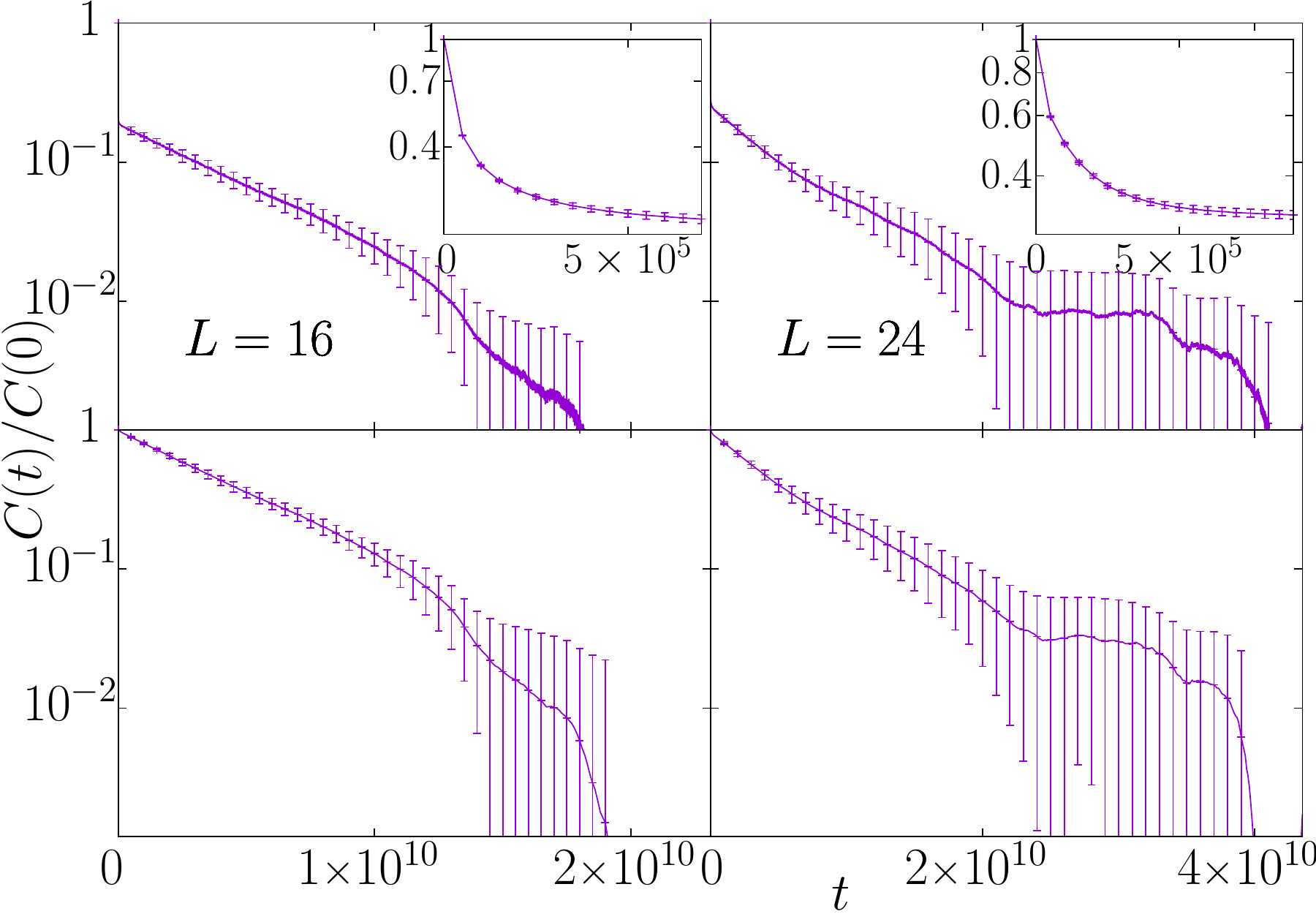}
\caption{\label{fig:func_corr} Auto-correlation function for the most
  chaotic sample for $L=16$ (left) and $L=24$ (right): (Top)
  Auto-correlation function computed using the method of
  \cite{janus:10} and (Bottom) using the variational method presented
  here.  {\bf Inset:} Linear-log plot showing the small $t$ behavior
  of the autocorrelation function.}
\end{figure}

An example of the improvement obtained in the computation of the
autocorrelation function is shown in Fig. \ref{fig:func_corr}. As it
can be inferred from Eq.~\eqref{eq:eigen-3}, a major difficulty is
that the amplitude for $\tau_\mathrm{exp}$, namely $A_{1,f}$, can be
very small. Indeed, the correlation function considered in a previous
work \cite{janus:10} (which is our piece-wise linear $f$, identifier
\#1 in Table \ref{tab:table_functions}, and $T^*$ set to the critical
temperature), has $A_{1,f}\approx 0.1$. Instead the optimized
autocorrelation function has an amplitude $A_{1,f}$ almost ten times
larger.

\begin{table}
\centering
\begin{tabular*}{0.8\columnwidth}{@{\extracolsep{\fill}}cc}
\br
Identifier & Function \\
\mr
\raggedleft
$0$ & piecewise constant \\
$1$ & piecewise linear\\
$2$ & piecewise quadratic\\
$3$ & piecewise cubic\\
$|$ & OR in couples\\
$\&$ & AND in couples\\
$\wedge$ &  XOR in couples\\
$*$ &  Multiplication in couples \\
\br
\end{tabular*}
\caption{\label{tab:table_functions} Different choices of the function
  $f$ used in the Variational Method.}
\end{table}

\begin{table}[ht!]
\centering
\begin{tabular*}{0.8\columnwidth}{@{\extracolsep{\fill}}cccccccccc}
\br
$L$ & $0$ & $1$ & $2$ & $3$ & $|$ & $\&$ & $\wedge$ & $*$ & Total  \\
\mr
$16$ & $2032$ & $5320$ &  $3875$ & $1374$  & $4$ & $115$ & $74$ & $6$ & $12800$ \\
$24$ & $1556$ & $7196$ &  $3089$ & $820$  & $0$ & $127$ & $11$ & $1$ & $12800$ \\
\br
\end{tabular*}

\caption{\label{tab:histo_func} Number of times the variational method
  has picked one of the eight choices among  the functions $f$ described in
  the text. $L$ denotes the lattice size.}
\end{table}

We observe in Table \ref{tab:histo_func} that, for almost all samples,
the variational method chooses a function $f$ depending on one system
copy only. Moreover, this variational method to a great extent
improves substantially the results obtained previously with a linear
function and a parameter $T^*$ chosen at the critical temperature
\cite{janus:10}.

\begin{figure}
\centering
\includegraphics[width=0.8\columnwidth]{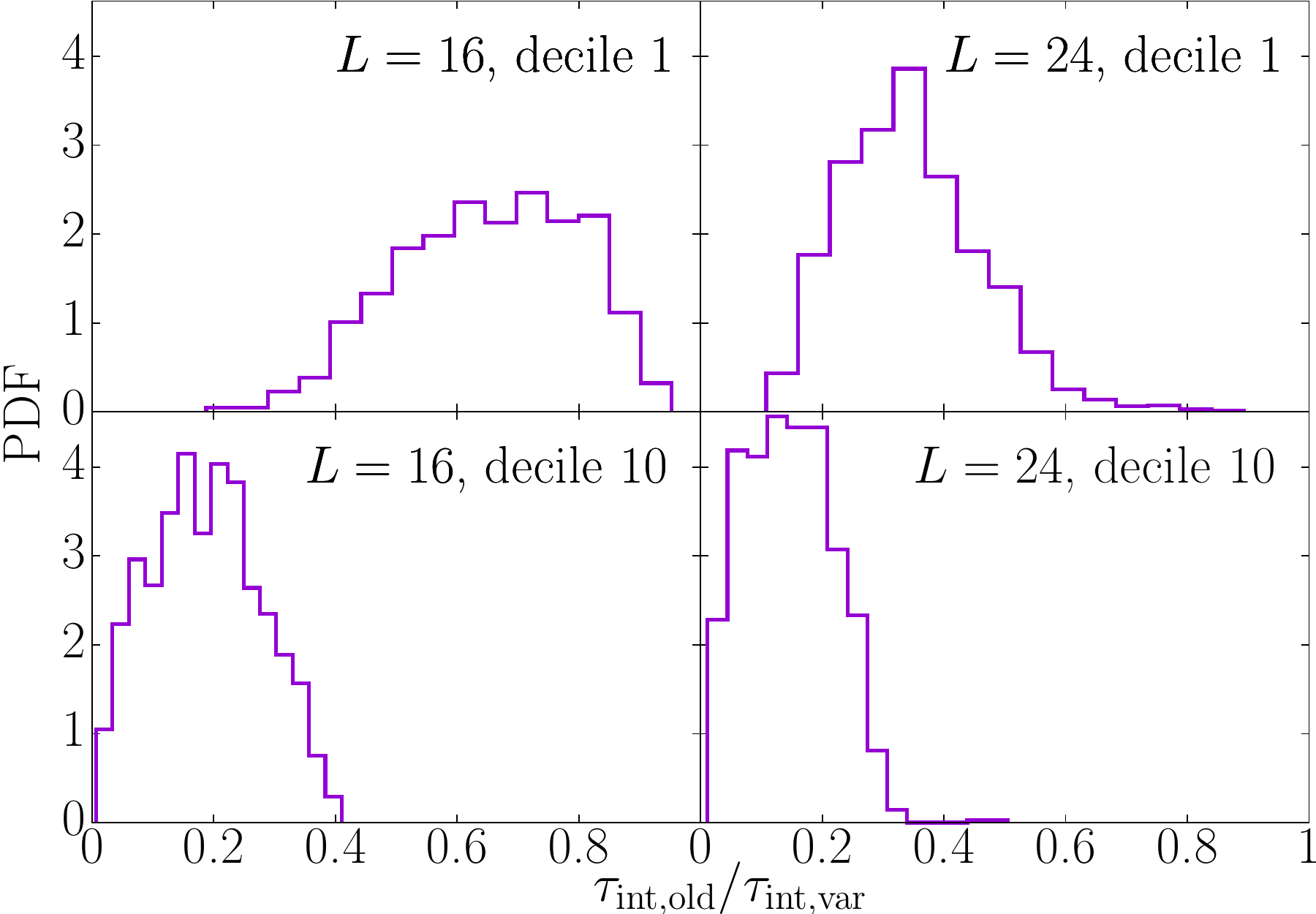}
\caption{\label{fig:histo_multiplot} Conditional probability density function of
  the ratio $\tau_{\mathrm{int,old}}/\tau_{\mathrm{int,var}}$, given
  that $\tau_{\mathrm{int,var}}$ belongs to a given decile. We show
  the data for the first decile (left) and the 10th-decile (right) for
  $L=16$ (top) and $L=24$ (bottom).}
 \end{figure}

We can do a quantitative comparison between the here proposed
variational method and the old approach. Let us histogram the ratio
$\tau_\mathrm{int,old}/\tau_\mathrm{int,var}$, conditioned to the value of
$\tau_\mathrm{int,var}$ (which is a good indicator of how chaotic a
sample is). We represent the result of this study in
Fig. \ref{fig:histo_multiplot} where
$\tau_\mathrm{int,old}/\tau_\mathrm{int,var}$ is represented for the first
and last deciles of $\tau_\mathrm{int,var}$\footnote{Deciles are
  similar to percentiles. First, samples are ordered according to
  their $\tau$. Then, we divide the samples in 10 sets (deciles) of
  equal size. Those samples with the lowest $\tau$ belong to decil 1,
  and so on.}.  The advantages of the variational estimator are evident
when one focus on decile 10 (i.e. for the most chaotic samples), where
we observe a significant fraction of samples with
$\tau_\mathrm{int,old}/\tau_\mathrm{int,var}<0.1$.

\subsection{The Finite Size Scaling Behavior of the Parallel Tempering Dynamics}\label{sect:dynamic_scaling}
In this section we study the Parallel Tempering dynamics for
$L=8,12,16,24$ and $32$, and we investigate temperature chaos from a
dynamical point of view.  In the following we will denote the mimimal
temperature allowed in a parallel tempering simulation as
$T_\mathrm{min}$ (it was $T_1$ in the previous section). We will focus
on the variational estimate of $\tau_\mathrm{int,var}$ (that we will
call simply $\tau$ from now on).

An implicit assumption of our study, corroborated by the results in
Sect.~\ref{sect:static-dynamic-chaos}, is that the scaling behavior of
$\tau$ is mostly decided by the value $T_\mathrm{min}$. Other details,
such as the number of temperatures in the parallel tempering mesh, are
expected to play a minor role (if kept in a reasonable range).  For
the comparative analysis of the dynamics we use the simulations at
$T_\mathrm{min}\approx 0.7$ shown in Table~\ref{tab:parametros_simu}.
This is the lowest value of $T$ we have thermalized for all our
lattice sizes. An important advantage of $T_\mathrm{min}\approx 0.7$
is that temperature-chaos has been already characterized at such
temperatures, in the equilibrium setting~\cite{fernandez:13}. Lowering
$T_\mathrm{min}$ would increase chaos effects, which would have been
good in principle, but it would have been also extremely difficult to
reach thermal equilibrium. Instead, increasing $T_\mathrm{min}$ to
approach the critical point would make the results irrelevant, because
samples displaying temperature chaos would be too scarce (besides, we
want to study the spin-glass phase, rather than critical effects).

For $L\leq 16$ we have $N_T=13$. For $L=24$ we needed to increase
$N_T$ in order to keep constant the acceptance rate of the temperature
exchange step of the parallel tempering simulation.  The $L=32$ data
are from Ref.~\cite{janus:10} and have been obtained with the
dedicated Janus computer~\cite{janus:09}. The Janus simulation used
heat bath dynamics, rather than Metropolis, and the Parallel Tempering
there had $N_T=34$ and $T_\mathrm{min}=0.703$. In order to be sure
that heat bath autocorrelation times are consistent with Metropolis
times (as we would expect) we simulated with Janus ten randomly
selected samples with both algorithms, finding that
$\tau_\mathrm{Metropolis}\approx \tau_\mathrm{heat-bath}/3$.

\begin{figure}
\centering
\includegraphics[width=0.8\columnwidth]{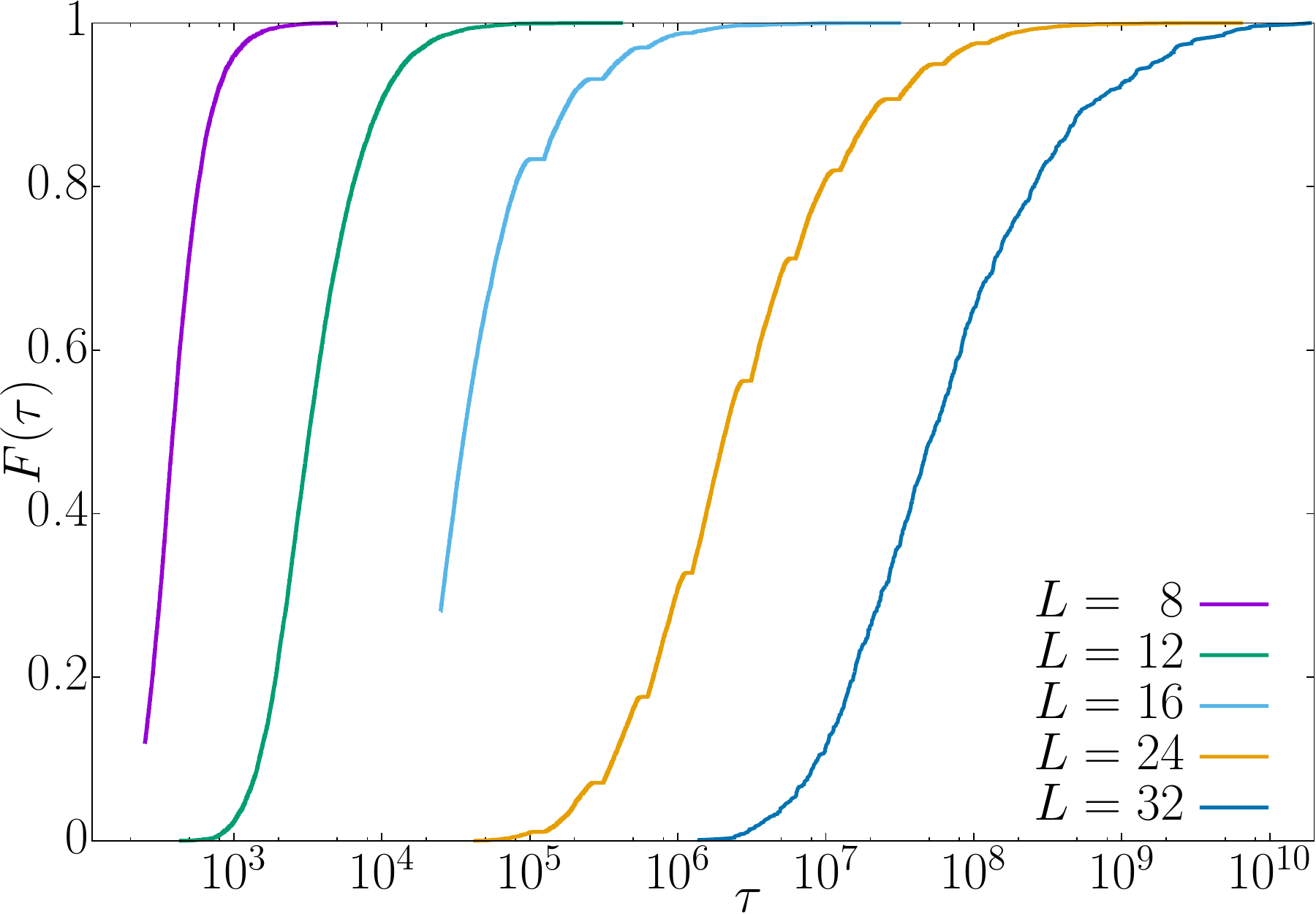}
\caption{\label{fig:all_L_prob_tau} Empirical probability distribution
  of $\tau$ for $L=8,12,16,24$ and $32$.  For $L=8$ and
  $L=12$ some of the samples have $\tau$ smaller than our minimal
  resolution (if $\tau<n_{\mathrm{Met}}$ we cannot compute it safely).
  We show only the part of the distribution function that can be
  safely computed.}
\end{figure}

We show in Fig.~\ref{fig:all_L_prob_tau} the cumulative distribution
function of $\tau$, $F(\tau)$.  The maximum slope of $F$ decreases
with $L$ for the small systems, and it stabilizes between $L=24$ and
$L=32$; indeed these two distributions can be approximately superposed
by a simple translation. This is reminiscent of a critical
slowing-down~\cite{zinn-justin:05}
\begin{equation}\label{eq:zPT-def}
\tau\sim L^{z^\mathrm{PT}(T_\mathrm{min})}\,.
\end{equation}
It is not obvious \emph{a priori} that such a simple scaling should
hold in the spin-glass phase. As a working, simplifying hypothesis we
assume that the exponent $z^\mathrm{PT}$ only depends on the value of
the lowest temperature in the parallel tempering grid,
$T_\mathrm{min}$ (and not on the number of temperatures).

\begin{figure}
\centering
\includegraphics[width=0.8\columnwidth]{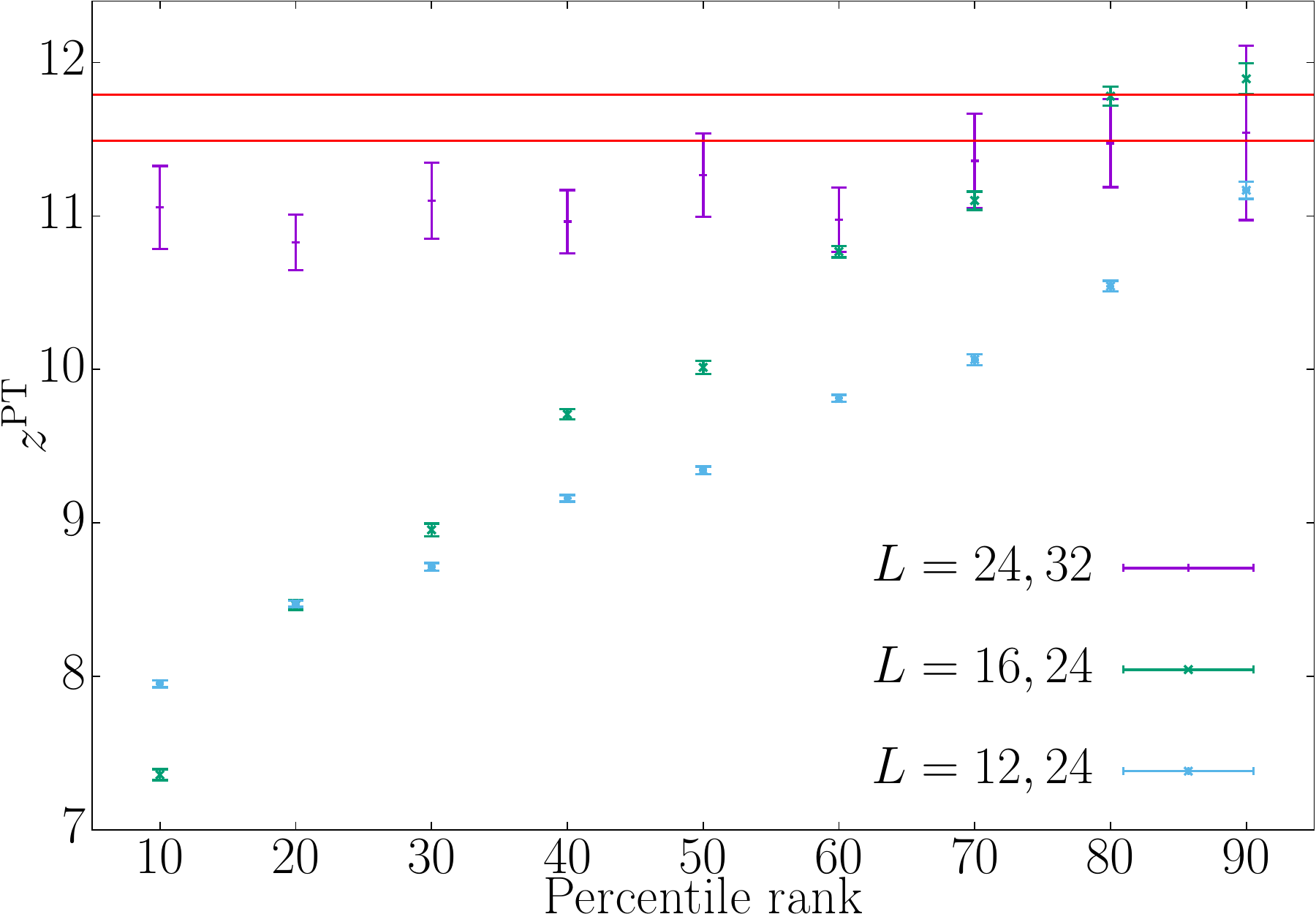}
\caption{\label{fig:z_percentile} The effective exponent
  $z^\mathrm{PT}(L_1,L_2,p)$ for three different pairs of lattice
  sizes (12,24), (16,24) and (24,32) as a function of the percentile
  rank $p$.  The two horizontal lines are the bounds for the
  off-equilibrium value $z=11.64(15)$ (see Eq.
  (\ref{eq:zeta_janus})). The numerical values of $z^\mathrm{PT}$ for
  the largest pair are compatible with the off-equilibrium value.}
\end{figure}

As a first test of Eq. (\ref{eq:zPT-def}) we compute an effective
$z$ exponent by comparing the probability distributions for two
lattice sizes $(L_1,L_2)$, by means of the definition
\begin{equation}
z^\mathrm{PT}(L_1,L_2,p)=\frac{\log(\tau(L_1,p)/\tau(L_2,p))}{\log(L_1/L_2)}\,,
\end{equation}
where $\tau(L_i,p)$ is determined by the implicit equation
$F(\tau(L_i,p))=p/100$ where $p=1,\ldots,100$ is the so called
percentile rank (i.e. $\tau(L_i,p)$ is the $p$-th percentile of the
distribution for the size $L_i$).  We have computed $z^\mathrm{PT}$
for three pairs of lattice sizes, (12,24), (16,24) and (24,32) and in
Fig. \ref{fig:z_percentile} we show the results as a function of the
rank.  The values for the largest pair, $(24,32)$, are independent of
the rank, within statistical errors, in agreement with the ansatz.
Smaller size couples give smaller estimates (in the same ball park)
for low ranks, but converge to the $(24,32)$ value for high ranks
(i.e. for the harder samples), and the coincidence improves and
extends to smaller ranks for larger lattices.  We remark that this
dynamic behavior is consistent with the static findings in this
temperature range~\cite{fernandez:13}: for $L=8$ it is almost
impossible to find samples displaying strong temperature chaos. One
needs to go to systems as large as $L=24,32$ to find chaotic samples with
a significant probability.

An interesting coincidence with the results of non-equilibrium
simulations~\cite{janus:08b,janus:09b,fernandez:15,janus:16} could
have a deep meaning.  Indeed in non-equilibrium conditions one finds
that the spin glass correlation length $\xi$, in a lattice of size
$L\gg\xi$, at temperature $T=0.7$ grows with the simulation time
$t_\mathrm{w}$ as~\cite{janus:09b}
\begin{equation}
\label{eq:zeta_janus}
\xi(t_\mathrm{w})\propto t_w^{1/z(T)}\,,\quad z(T=0.7)=11.64(15)\,,
\end{equation}
where $z(T)$ is the so-called dynamic critical exponent, that turns
out to be strongly temperature dependent in the spin-glass phase
$z(T)\propto T_\mathrm{c}/T$.
Our results for the lattice pair
$(24,32)$ suggest that
\begin{equation}\label{eq:z-coincide}
z(T=0.7)\approx z^\mathrm{PT}(T_\mathrm{min}=0.7)\,. 
\end{equation}

\begin{figure}
\centering
%\hspace{-1cm}
\includegraphics[width=0.8\columnwidth]{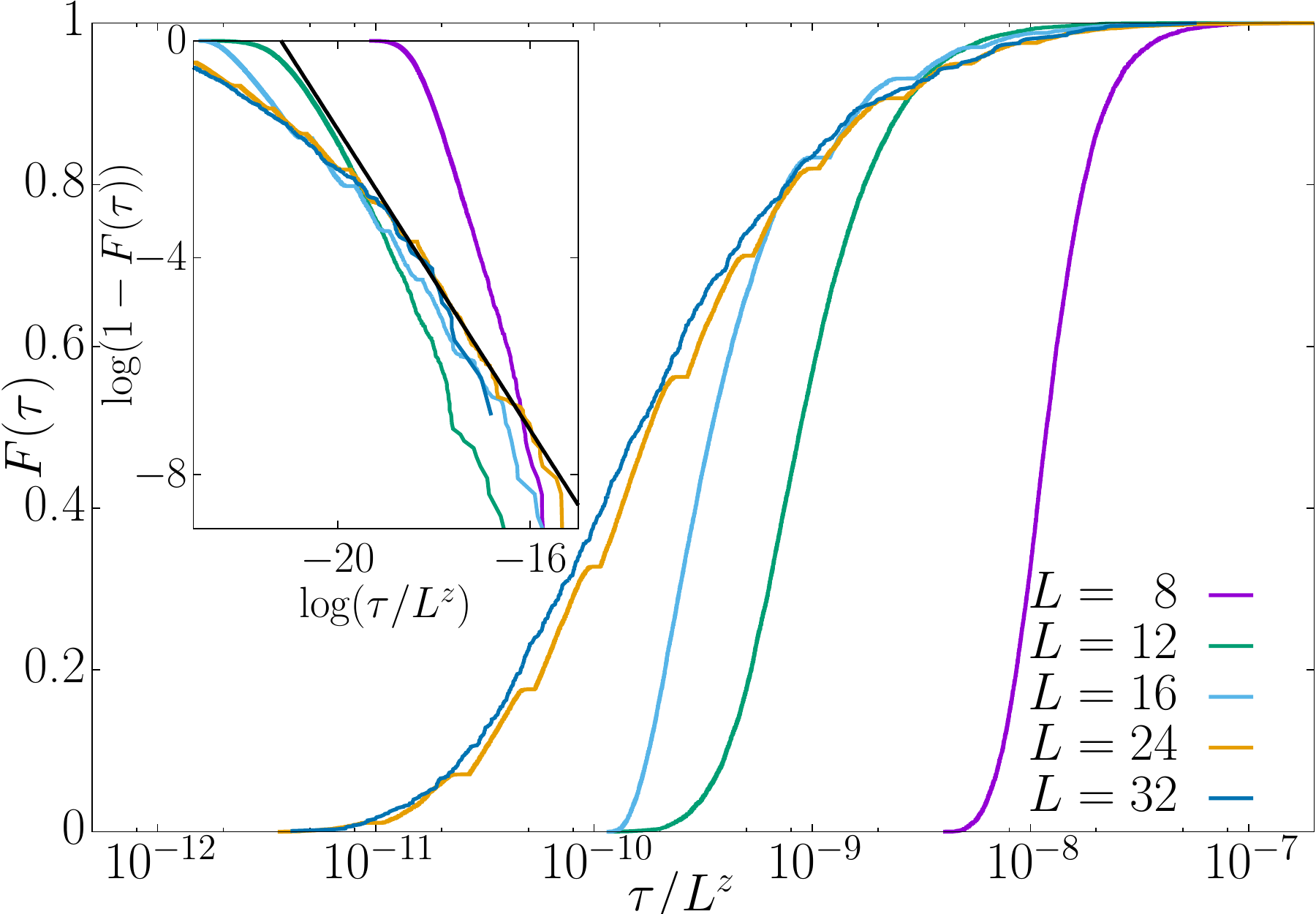}
\caption{\label{fig:all_L_prob_tau_reescaled} Probability distribution
  function of the rescaled variable $y = \tau/L^{z}$, [$z$ is the
    dynamic exponent corresponding to $T_\mathrm{min}=0.7$, namely
    $z(T=0.7)=11.64(15)$]. {\bf (Inset)} Plot of $\log(1-F(\tau))$
  versus $\log\left(\tau/L^{z}\right)$; the straight black line is a
  fit to the form $a_0 - a_1 \log(\tau/L^z)$ yielding $a_0 = -29.33$
  and $a_1=-1.38$.}
\end{figure}

As a further test we can rescale the whole probability distribution
by using Eqs.~\eqref{eq:zPT-def} and~\eqref{eq:z-coincide}.  This is
done in figure \ref{fig:all_L_prob_tau_reescaled} (main) that shows
$F(\tau)$ as a function of $y = \tau/L^z$.  As expected, the data for
$L=24$ and $L=32$ present a nice collapse.  The curve corresponding to
$L=16$ collapses with them for percentile ranks higher than $80$ only
and the curve corresponding to $L=12$ collapses for percentile
ranks higher than $90$.
This is a nice and smooth behavior. On the larger lattice sizes we
reach a perfect scaling, but already on smaller lattice we see a
partial scaling, that improves for increasing size. 
In figure~\ref{fig:all_L_prob_tau_reescaled}
(inset), we show a log-log plot of $1\!-\!F(\tau)$ as a function of
$\tau/L^z$, that emphasizes the large $\tau$ tail of the
distribution. The fit presented shows that the probability density
function of $\tau$ behaves, asymptotically for large $y$, like a
fat tailed distribution:
\begin{equation}
\rho(y\equiv \tau/L^z) \sim y^{-1-a_1}\,,\quad a_1\approx 1.38 \,.
\end{equation}
The distribution seems to reach its
asymptotic form for $L\geq 24$. Perhaps unsurprisingly, the
thermodynamic (i.e. equilibrium) effective potential that
characterizes temperature chaos turns out to be also asymptotic for
$L\geq 24$~\cite{fernandez:13}.

\begin{figure}
\centering
\includegraphics[width=0.8\columnwidth]{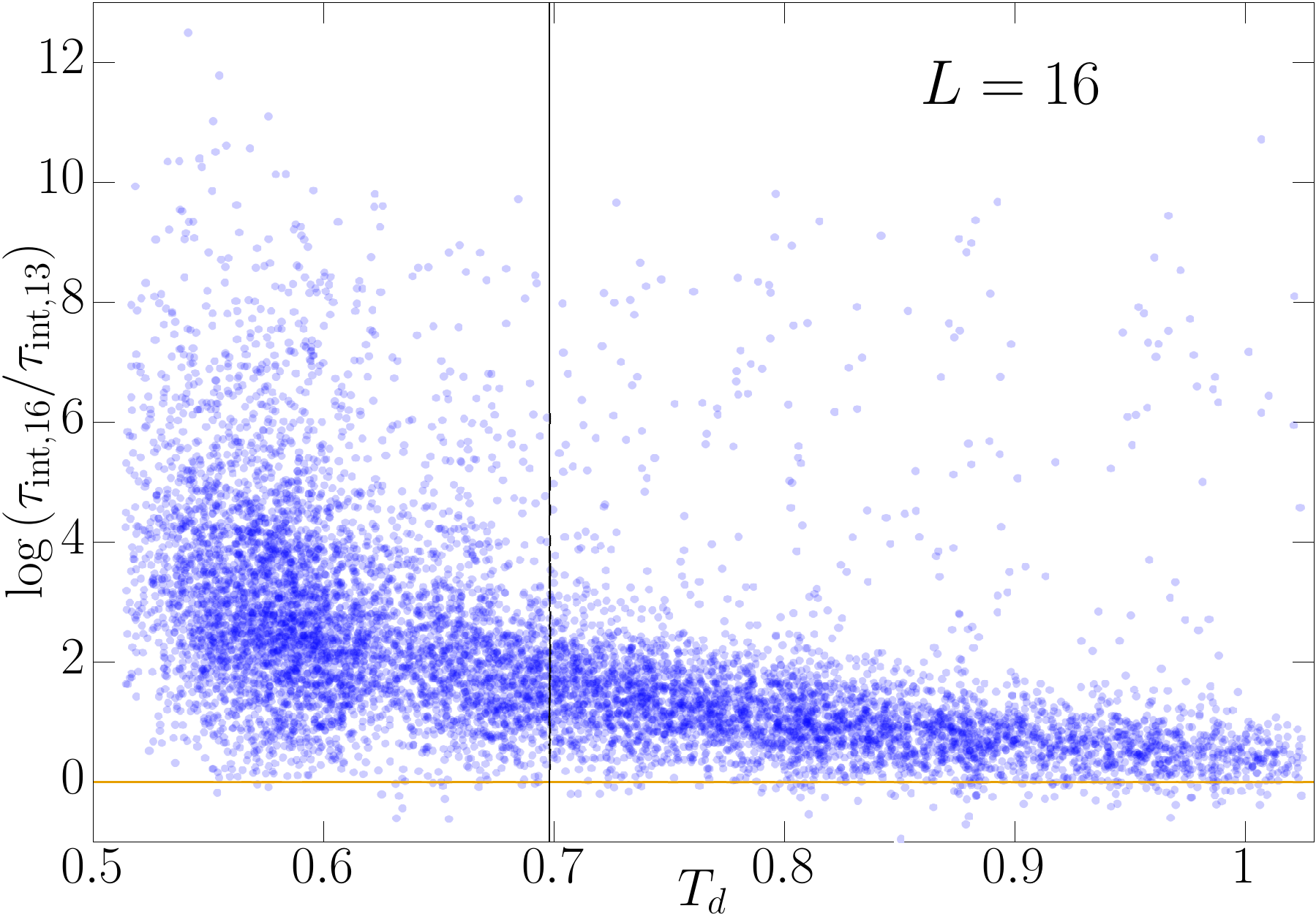}
\caption{\label{fig:cociente_taus} Scatter plot of
  $\log(\tau_{\mathrm{int},16}/\tau_{\mathrm{int},13})$ versus $T_d$.
  The lattice size is $L=16$, $\tau_{\mathrm{int},16}$ is the
  relaxation time for $N_T=16$ ($T_\mathrm{min}=0.479$),
  $\tau_{\mathrm{int},13}$ is the relaxation time for $N_T=13$
  ($T_\mathrm{min}=0.698$), $T_d$ is the temperature of chaos from a
  dynamical point of view (defined in the variational method) of the
  simulation with $N_T = 16$. Same disorder samples in the two
  simulations. The vertical black line represents the minimum
  temperature simulated in the $N_T= 13$ simulation. (We added a small
  Gaussian white noise to $T_d$, which is a discrete variable, to
  avoid the cluttering of data in vertical lines). }
\end{figure}

In order to study how the range of temperatures in the parallel
tempering affects the dynamics, we have performed an extra simulation
for $L=16$. In the new simulation we take a lower minimum temperature
($T_\mathrm{min}=0.479$ instead of $T_\mathrm{min}=0.698$) increasing
$N_T$ from $13$ to $16$ in order to keep the interval between adjacent
temperatures fixed, see Table~\ref{tab:parametros_simu}.  Since the
simulation with $N_T = 16$ reaches a lower minimum temperature than
the simulation with $N_T = 13$ we expect to find chaos events (i.e a
jam in the parallel tempering temperature flow) that the simulation
with $N_T=13$ cannot see.
In Fig.~\ref{fig:cociente_taus} we show a scatter plot of
$\log(\tau_{\mathrm{int},16}/\tau_{\mathrm{int},13})$ versus $T_d$ for
the 12800 samples ($\tau_{\mathrm{int},16}$ and
$\tau_{\mathrm{int},13}$ are the autocorrelation times for $N_T=16$
and 13 respectively. $T_d$ is the temperature $T^*$ where the
variational estimate $\tau_{\mathrm{int},f}$ reaches its maximum).

For $T_d>0.698$  the ratio takes
values of order one for most samples, while for $T_d < 0.698$ there is a huge number of
samples with $\tau_{\mathrm{int},16} \gg \tau_{\mathrm{int},13}$,
i.e. there are a lot of samples with a chaotic behavior in a temperature-range 
below $T_\mathrm{min}=0.698$.

\begin{figure}
\centering
\includegraphics[width=0.8\columnwidth]{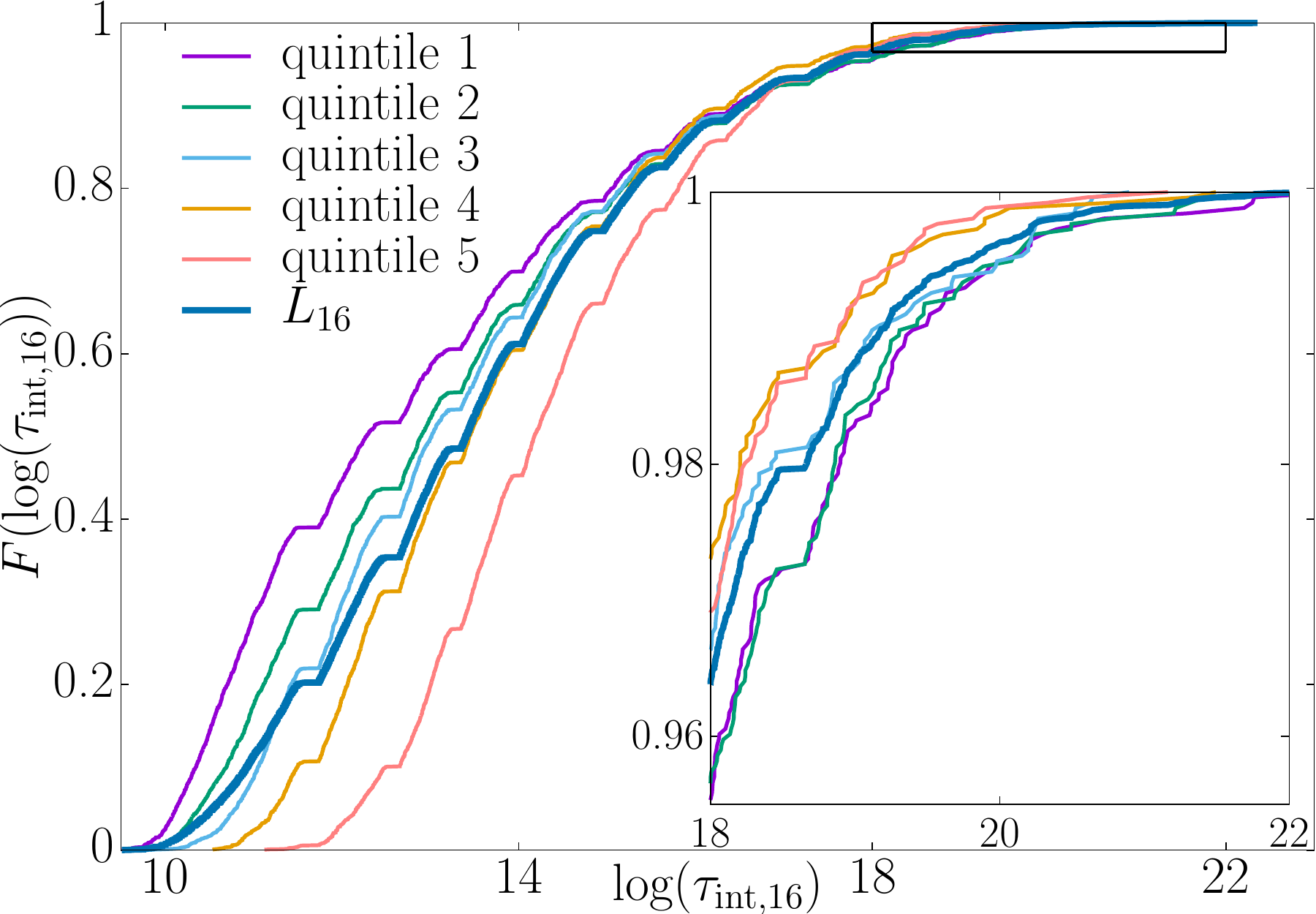}
\caption{\label{fig:no-prediction-from-small-Tmin}
  The empirical probability distribution
  as a function of $\tau$ for the $N_T=16$ simulation, conditional to the
  $\tau$ obtained from $N_T=13$ simulation belonging to a given quintile.  The
  non-conditional probability distribution function is
  also shown ($L_{16}$ curve).  {\bf Inset.} Blowup of the top right
  part of the main figure. For the
  hard samples, the simulation with $T_\mathrm{min}=0.698$
  conveys little or no information on the difficulty of the
  $T_\mathrm{min}=0.479$ simulation.}
\end{figure}

The same idea can be analyzed from a different point of view. Imagine
that we have studied with great care a given sample down to some
temperature $T_\mathrm{min}$. Can we say something about possible
chaotic effects at lower temperatures? The question is answered
negatively in Fig.~\ref{fig:no-prediction-from-small-Tmin}: the
probability that a sample has a large $\tau_{\mathrm{int}}$ for the
simulation with a lower $T_\mathrm{min}$ is not correlated to the
value of $\tau_{\mathrm{int}}$ for the first simulation.

%%%%%%%%%%%%%%%%%%%%%%%%%%%%%%%%%%%%%%%%%%%%%%%%%%%%%%%%%%%%%%%%%%%%%
\subsection{Statics}\label{sect:static-chaos}

In the infinite volume limit, static temperature chaos is the complete
rearrangement of the equilibrium configuration under any change of
temperature.  It has been studied numerically mostly through the
disorder average of the probability density function of the overlap
between the spin configurations at temperatures $T_1$ and $T_2$

\begin{equation}
q_{T_1,T_2} = \frac{1}{V} \sum_x s_x^{T_1} s_x^{T_2} \, ,
\end{equation}
or through ratio of moments of this distribution.  However, because of
the size of the systems that can be currently simulated, the overlap
is strongly influenced by finite size effects. It has been suggested
that static temperature chaos is a rare events driven phenomena, that
should be studied via the distribution of the sample-dependent chaotic
parameter \cite{fernandez:13,billoire:14}:
\begin{equation}
X_{T_1,T_2}^J =  \frac{\braket{q^2_{T_1,T_2}}_J}{\sqrt{\braket{q^2_{T_1,T_1}}_J \braket{q^2_{T_2,T_2}}_J}} \, ,
\end{equation}
where $\braket{\cdot\cdot \cdot}_J$ is the thermal average within a
given sample ($J$). Notice that $0 < X^{J}_{T_1,T_2} \lesssim 1$;
$X^J_{T_1,T_2} = 1$ means that equilibrium spin configuration of the
$J$ sample at temperature $T_1$ and temperature $T_2$ are
indistinguishable while $X^J_{T_1,T_2} = 0$ means that the equilibrium
spin configurations are completely different.

The temperature evolution of $X^{J}_{T_1,T_2}$ for some selected
samples is shown in Fig.~\ref{fig:xvsT} (in the figure $T_1$ is kept
fixed to $T_1=T_\mathrm{min}$ while $T_2$ is made to vary). In some
samples, we find \emph{chaotic events}, namely sharp drops of
$X^{J}_{T_1,T_2}$ at very well defined temperatures, implying that
the typical spin configurations significantly differ at the two sides
of the chaotic event. It was empirically observed in
Ref.~\cite{fernandez:13} that chaotic events occurring at low
temperatures are most harmful to the performance of parallel
tempering. To quantify the effect, 
 the chaotic integral $I$ was introduced
\begin{equation}
I = \int_{T_{\mathrm{min}}}^{T_{\max}} X^J_{T_{\mathrm{min}},T_2}\ \mathrm{d}T_2 \, .
\end{equation}
Note that a sharp drop of $X^J_{T_{\mathrm{min}},T_2}$ at a low $T_2$ will result into
a very low value of the chaotic integral $I$. Furthermore,
a study of the temperature behavior of the chaotic parameter leads to
the conclusion that chaos events happen at low temperatures only,
therefore the high temperatures introduce only noise in the estimate
of $I$. In order to eliminate this noise we introduce a new integrated
chaotic parameter $I_2$ that involves the half-lower temperatures
only.

Nevertheless, there exists certain samples that exhibit a huge
$\tau_{\mathrm{int}}$ and have a relative large chaotic integral, so the
correlation between static and dynamics is more complicated than one 
could hope. Therefore, in order to improve our thermodynamic understanding of 
the Parallel Tempering dynamics, we need to look elsewhere. We have found
it useful to consider the temperature derivative of the chaotic parameter.
Indeed, it is easy to prove
that:
\begin{equation}
\left. dX^J_{T_{1},T_2}/dT_2 \right|_{T_2=T_1} = 0 \, .
\end{equation}
for any temperature $T_1$.  However, if we focus on these outlier
samples, we notice that these samples present a sharp drop in
$X_{T_1,T_2}^J$ at two consecutive temperatures.  This
observation will motivate the definition in Eq.~\eqref{eq:K-def}, below.

\subsection{Correlations Dynamics-Static}\label{sect:static-dynamic-chaos}

Once we have characterized the chaos phenomena from both dynamical and
static point of view, we are interested in knowing how these static
and dynamics estimators are correlated. 

Besides the chaos integrals $I$ and
$I_2$, we introduce a new
quantity for further use:
\begin{equation}\label{eq:K-def}
K_i = 1-X^J_{T_i,T_{i+1}}
\end{equation}
After some trials, we have finally defined a last parameter:
\begin{equation}
I_X = a I_2 - b \min_i \left(-\log \left(K_i^2\right)\right) - c\sum_i \left(-\log \left(K_i^2\right)\right)  \, ,
\label{eq:IX}
\end{equation}
where the coefficients $a$, $b$ and $c$, that depends on the lattice
size, are obtained through a minimization of the correlation
coefficient $r$ between
  $I_X$ and $\log(\tau_\mathrm{int})$
($r$ is negative, and it would be $r=-1$
if we managed to achieve a perfect understanding of our dynamical
data). The values of these coefficients are given in Table
\ref{tab:coef_adjust}

\begin{table}[ht] 
\centering
\begin{tabular*}{0.8\columnwidth}{@{\extracolsep{\fill}}cccc}
\br
$L$ & $a$ & $b$ & $c$  \\
\mr
$16$ & $0.6143$ & $0.2865$ & $0.1373$ \\
$24$ & $0.2963$ & $0.3217$ & $0.0120$ \\
\br
\end{tabular*}

\caption{\label{tab:coef_adjust} Value of the coefficients $a$, $b$
  and $c$ in Eq. \ref{eq:IX}, that maximize the correlation between
  $I_X$ and $\log(\tau_\mathrm{int})$.}
\end{table}
 
\begin{figure}[ht!]
\centering
\includegraphics[width=0.8\columnwidth]{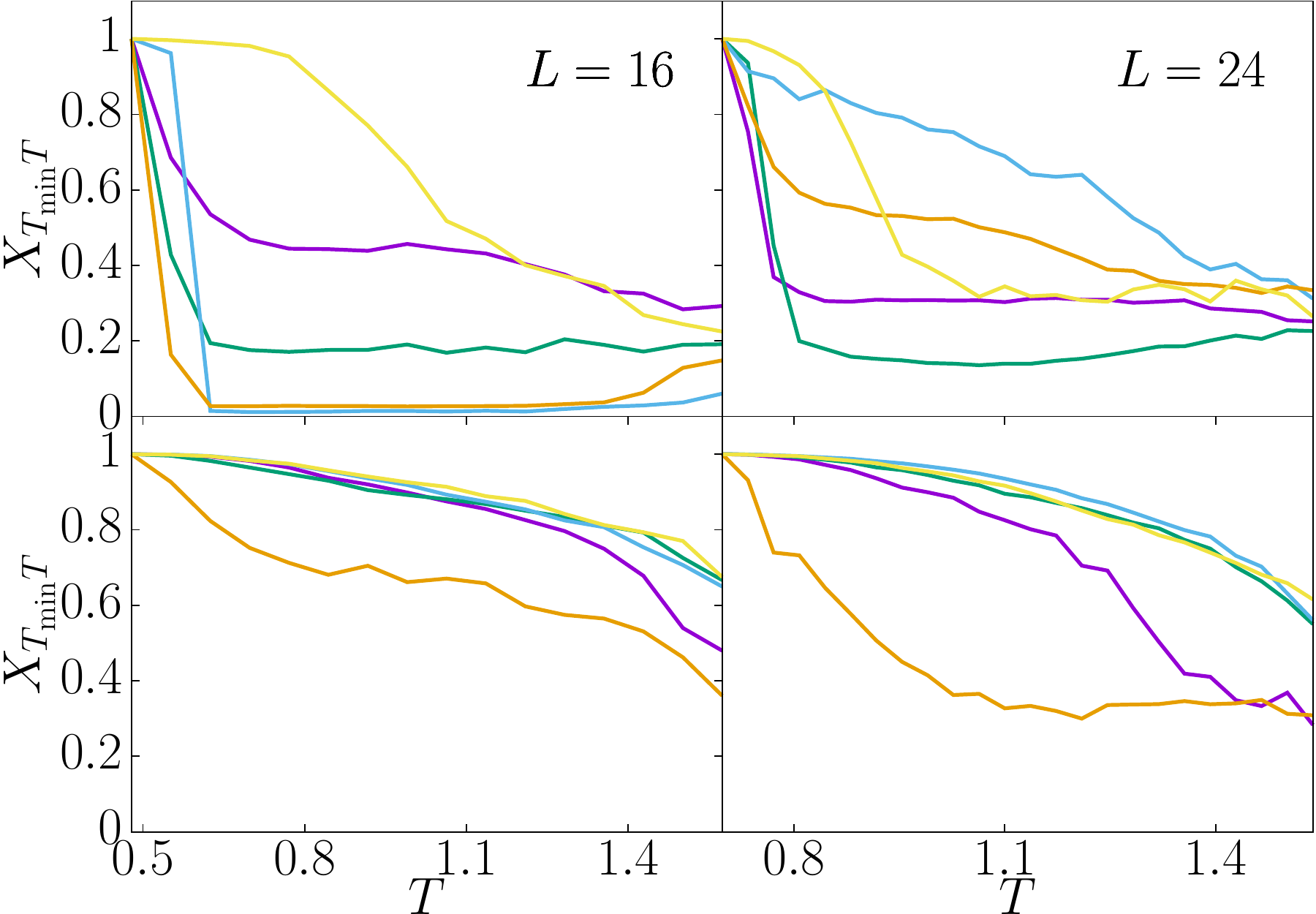}
\caption{\label{fig:xvsT}  Plot of $X_{T_{\mathrm{min}},T}^J$ versus
  $T$ for the five most chaotic samples (top) and the five less
  chaotic ones (bottom):  $L=16$ case (left) and
   $L=24$ case (right).}
\end{figure}

This finding is supported by Fig. \ref{fig:xvsT}. We see that the most
chaotic samples in terms of the integrated autocorrelation time
(Fig. \ref{fig:xvsT}, top), present a sharp fall in the chaotic
parameter. On the other hand, we can see that less chaotic samples in
terms of the integrated time (Fig. \ref{fig:xvsT}, bottom), have a
much smoother fall.

\begin{figure}[ht]
%\hspace{-0.5cm}
\centering
\includegraphics[width=0.8\columnwidth]{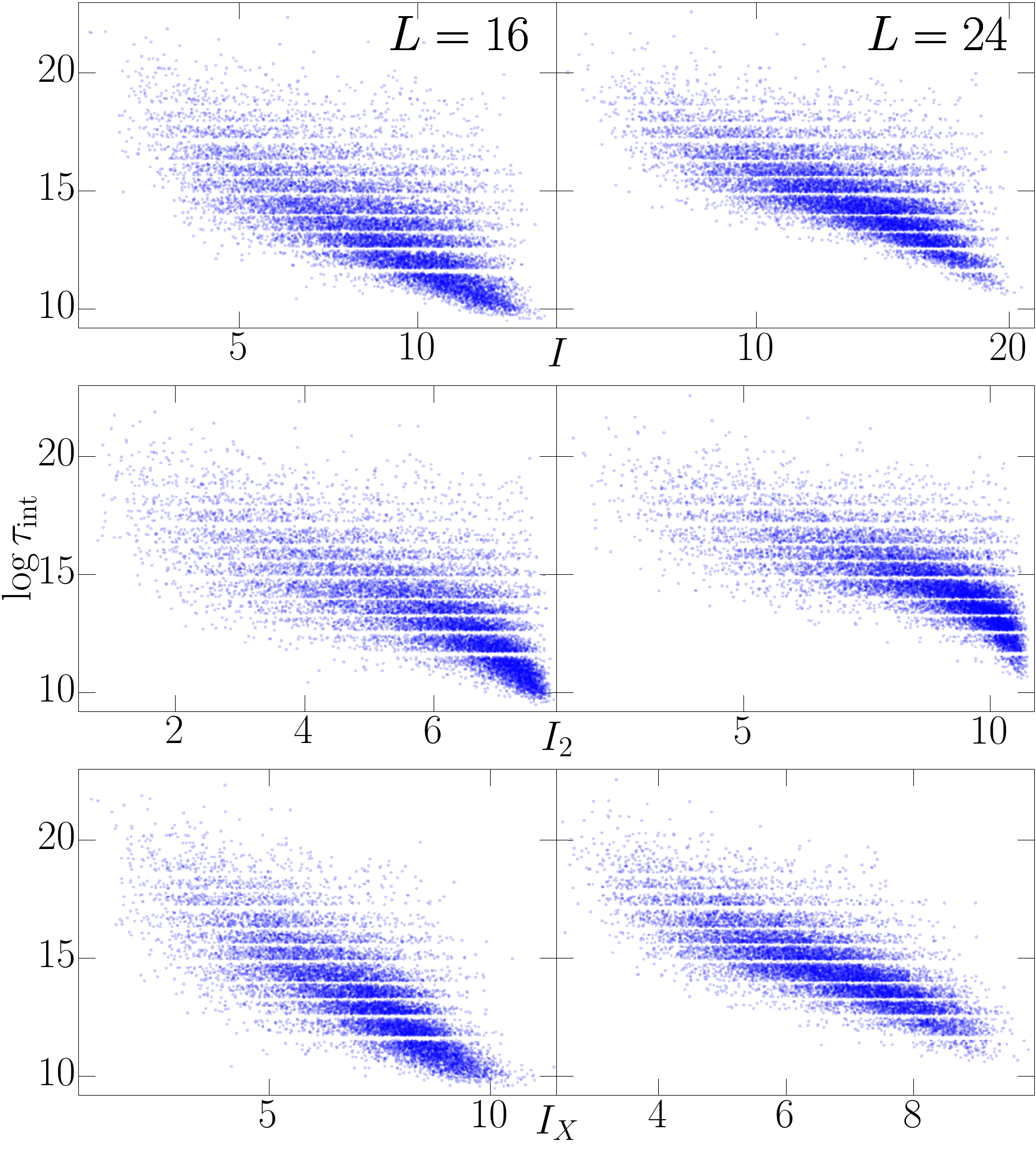}
\caption{\label{fig:log_tau_I} Scatter plot of
  $\log(\tau_{\mathrm{int,var}})$ versus the integrated chaotic
  parameter $I$. We present data for two lattice sizes and for the
  three definitions of the integrated chaotic parameter defined in the
  text ($I, I_2$ and $I_X$). The pattern of depleted horizontal
  bands is due to our choice of a few $l_\mathrm{blo}$.}
\end{figure}

In Fig. \ref{fig:log_tau_I}, we confront the most representative
estimator for the dynamical chaos, namely the largest integrated
autocorrelation time $\tau_\mathrm{int}$ found in our variational study,
with the static chaotic integrals $I$, $I_2$ and $I_X$. We can observe
how spurious values of the original parameter $I$ (i.e. large values
of $I$ associated to large $\tau_\mathrm{int}$) are displaced towards
lower values when we use the improved parameters $I_2$ and $I_X$.

The  value of the correlation
coefficients are reported in Table \ref{tab:coef_correla}.\footnote{
Statistical-error estimates were
computed using the bootstrap method.} We observe a
strong anti-correlation in  $I_X$, that
improves over the previous indicator of correlation
$I$. \cite{fernandez:13} The improvement is less clear for $I_2$.

\begin{table}[ht]
\centering
\begin{tabular*}{0.8\columnwidth}{@{\extracolsep{\fill}}ccccc}
\br
&$L$ & Integral & $r$  \\
\mr
&$16$ & $I$   & $-0.714 \pm 0.005$ \\
&$16$ & $I_2$ & $-0.751 \pm 0.005$ \\
&$16$ & $I_X$ & $-0.795 \pm 0.004$ \\
\br
&$24$ & $I$   & $-0.725 \pm 0.005$ \\
&$24$ & $I_2$ & $-0.746 \pm 0.005$ \\
&$24$ & $I_X$ & $-0.786 \pm 0.004$ \\
\br
\end{tabular*}
\caption{\label{tab:coef_correla} Correlation coefficients for the
  scatter plot of $\log(\tau_{\mathrm{int}})$ versus the integrated
  chaotic parameter, for each two lattice size and for the three
  definitions of the parameter ($I, I_2$ and $I_X$).}
\end{table}

\begin{figure}[ht]
\centering
\includegraphics[width=0.8\columnwidth]{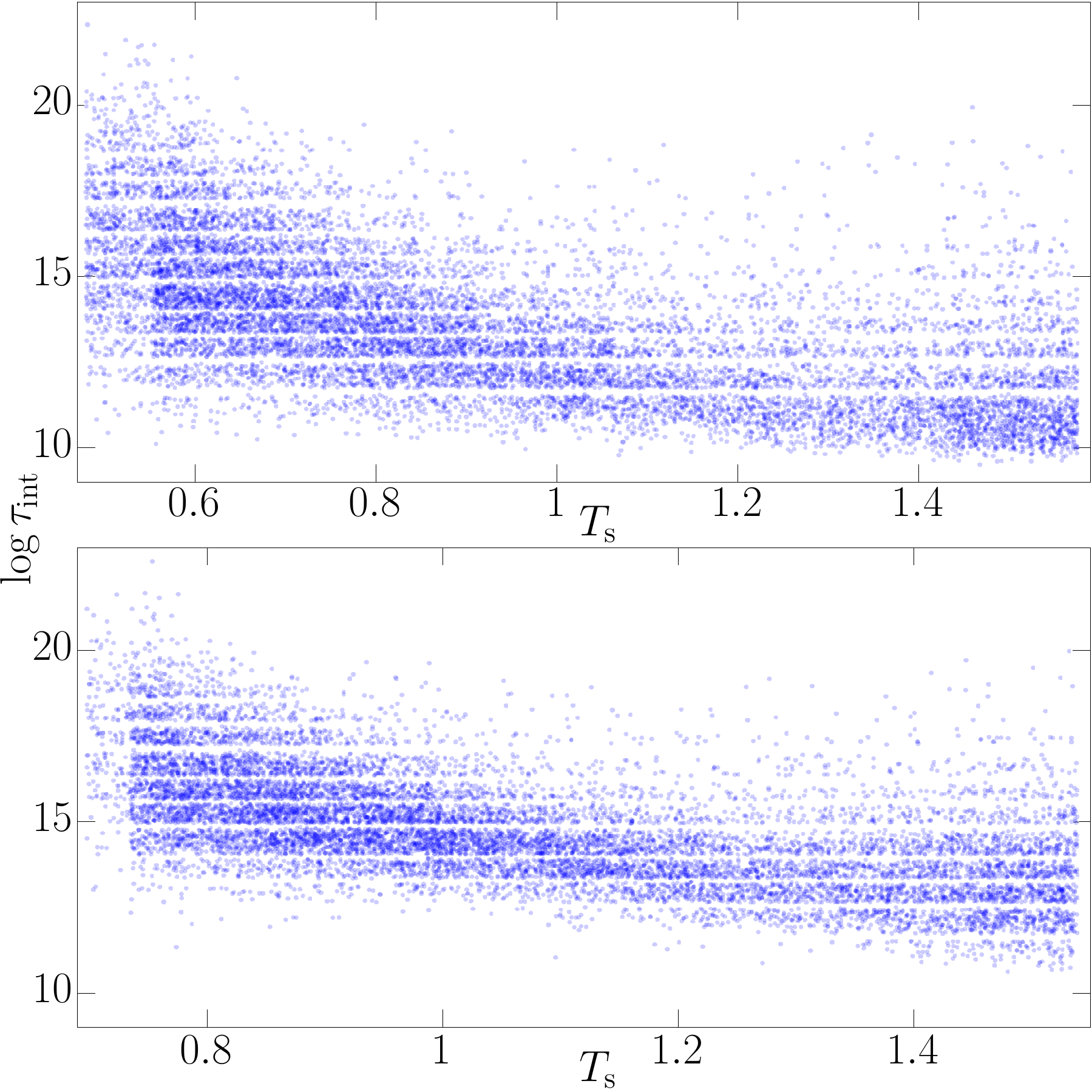}
\caption{\label{fig:log_tau_Ts} Scatter plot of
  $\log(\tau_{\mathrm{int}})$ against $T_\mathrm{s}$. We show $L = 16$
  (top) and $L=24$ (bottom).  $T_s$ is the temperature where
  $X_{T_{\mathrm{min}},T}^J$ presents the maximum (negative) slope.  }
\end{figure}

We can try to define other magnitudes (wither static or dynamical)
that capture the chaos phenomenon.  One possible choice is the
temperature, $T_s$, in which $X_{T_{\mathrm{min}},T}^J$ presents the
maximum (negative) slope. Unfortunately, we observe a weaker
correlation between both estimators, $\tau$ and $T_s$, (see
Fig. \ref{fig:log_tau_Ts}) and we can check it quantitatively through
Table \ref{tab:coef_correla_Ts}. Some further attempts along these
lines are explored in~\ref{app:quatities_not_chaos}.

\begin{table}[ht]
\centering
\begin{tabular*}{0.8\columnwidth}{@{\extracolsep{\fill}}ccccc}
\br
&$L$   && $r$ & \\
\mr
&$16$  && $-0.621 \pm 0.006$ &\\

&$24$  && $ -0.621 \pm 0.006$ &\\
\br
\end{tabular*}
\caption{\label{tab:coef_correla_Ts} Correlation coefficients for the
  scatter plot of $\log(\tau_{\mathrm{int}})$ versus $T_\mathrm{s}$
  for the two simulated lattice sizes.}
\end{table}

\section{Discussion and conclusions}\label{sect:conclusions}

We have proposed an efficient variational method to estimate the
elusive exponential autocorrelation time of a Monte Carlo Markov
chain, specific to the (arguably important) case of a Parallel Tempering
simulation. In this variational method we have introduced three parameters
(a temperature $T^*$, a function $f$ and a block length). We have
checked that this procedure is very robust and can easily be
implemented in an automatic way.

In addition, we have studied the scaling properties of the probability
distribution of the autocorrelation time, obtained using the proposed
variational approach. In particular we have shown that scaling holds
for lattices of sizes $L\ge 24$, consistently with previous studies
using effective potentials.

Moreover, we have introduced additional static chaotic indicators, and
finally we have checked the statistical correlations between these
static chaotic indicators and the dynamical correlation times.

\section{Acknowledgments}

We thank the Janus collaboration for allowing us to analyze the $L=32$
autocorrelation times from Ref.~\cite{janus:10}. We also are grateful
for allowing us to carry out a short simulation on the Janus computer,
in order to establish the correspondence between the Metropolis and
the Heat-bath autocorrelation times.

This project has received funding from the European Research Council
(ERC) under the European Union’s Horizon 2020 research and innovation
program (grant agreement No 694925). We were partially supported by
MINECO (Spain) through Grant Nos. FIS2015-65078-C2, FIS2016-76359-P
and by the Junta de Extremadura (Spain) through Grant No. GRU10158
(these three contracts were partially funded by FEDER). Our simulations were
carried out at the BIFI supercomputing center (using the
\emph{Memento} and \emph{Cierzo} clusters), at the TGCC supercomputing
center in Bruy\`eres-le-Ch\^atel (using the \emph{Curie} computer,
under the allocation 2015-056870 made by GENCI) and at ICCAEx
supercomputer center in Badajoz (\emph{grinfishpc} and
\emph{iccaexhpc}). We thank the staff at BIFI, TGCC and ICCAEx
supercomputing centers for their assistance.

\appendix
\section{Parameters of the simulation}
\label{app:simulation_parameters}
Whereas in numerical simulations of spin glasses the disorder samples
are usually independent, the samples we use here are not fully
independent.  The motivations of our choice are explained in
Ref. \cite{billoire:17}.  We consider cubes with $L^3$ spins and
$3L^3$ couplings, divided into an inner part of $(L/2)^3$ spins and an
outer part surrounding it. We simulate $10$ independent inner samples,
and, for each inner sample, $1280$ independent outer samples. We
simulate four replica (independent spin systems) for every inner and
outer sample.  Hence we have simulated $12800$ disorder realizations
(samples) with a total of $12800\times 4$ real spin systems.  The
parameters of the simulation can be found in Table
\ref{tab:parametros_simu}.

The thermalization criteria that have been used is as follows (as
explained above these criteria applied to every sample, individually). First of
all, the number of iterations in $\tau_\mathrm{exp}$ units
($l_\mathrm{blo} = 1$) must be greater than $20$; as a double-check to
avoid failures in the automated fitting procedure, we recomputed
$\tau_\mathrm{exp}$ with $l_\mathrm{blo} = 10$ (the total simulation
length is also required to be longer than
$20\tau_\mathrm{exp}^{l_\mathrm{blo}=10}$). 

However, we had some additional safety checks to ensure that the
computation of $\tau_\mathrm{exp}$ could be trusted. For those samples
where any of the following two requirements was not met, we doubled
the total simulation length and, only afterwards, we recomputed
$\tau_\mathrm{exp}$. First, in order to make sure that every sample spends
enough time at high temperatures, we require that each copy of the
system in the Parallel Tempering method spends at least $35\%$ of the
time in the upper half temperature region. Second, the ratio between the
larger and the smaller values of $\tau_{\mathrm{int}}$, as computed for
each of the four independent replica, must be less than two (for
either $l_\mathrm{blo} = 1,10,100$).  This last requirement can help us
to identify a lack of thermalization for those samples whose leading
term in the autocorrelation function has a very small amplitude.

\begin{table}
\centering
\begin{tabular*}{0.9\columnwidth}{@{\extracolsep{\fill}}cccccccc}           

\multicolumn{7}{c}{MUSA-MSC} \\
\br
$L$ & $L_{\mathrm{int}}$ & $N_T$ & $T_{\mathrm{min}}$ & $T_{\max}$ & $N_\mathrm{Met}$ ($ \times 10^6$) & $\mathrm{ps/s}$  \\
\mr
24 & 12 & 24 & 0.698 & 1.538 & 500 & 104 \\ 
16 & 8 & 16 & 0.479 & 1.575 & 250 & 107  \\ 
16 & 8 & 13 & 0.698 & 1.575 & 250 & 119  \\ 
16 & 12 & 13 & 0.698 & 1.575 & 250 & 119 \\ 
14 & 12 & 13 & 0.698 & 1.575 & 500 & 120 \\
12 & 6 & 13 & 0.698 & 1.575 & 250 & 119  \\ 
8 & 4 & 13 & 0.698 & 1.575 & 250 & 126 \\
\br
\end{tabular*}
\\[2mm]

\begin{tabular*}{0.9\columnwidth}{@{\extracolsep{\fill}}ccccccccc}           

\multicolumn{8}{c}{MUSI-MSC} \\
\br
$L$ & $L_{\mathrm{int}}$ & $N_T$ & $N_\mathrm{samp}$ & $N_\mathrm{Met,min}$ & $N_\mathrm{Met,mean}$ & $N_\mathrm{Met,max}$ & $\mathrm{ps/s}$ \\
\multicolumn{4}{c}{} & $\times 10^6$ & $\times 10^6$ & $\times 10^6$ & \\
\mr
24 & 12 & 24 & 2441 & 1000 & 4262 & 326000 & 57 \\ 
16 & 8 & 16  & 2898 & 500 & 5096 & 355500 & 304 \\ 
16 & 8 & 13 & 338 & 500 & 543 & 4000 & 306 \\ 
16 & 12 & 13 & 314  & 500 & 578 & 8000 & 306 \\ 
%14 & 12 & 13 & 0 & -  & - & - & - \\
%12 & 6 & 13 & 0 & -  & - & - & - \\ 
%8 & 4 & 13 & 0 & -  & - & - & -
\br
\end{tabular*}

\caption{\label{tab:parametros_simu} Parameters of the
  simulations. $L$ is the lattice size; $L_\mathrm{int}$ the size of the
  inner part of the lattice; $N_T$, $T_{\mathrm{min}}$ and $T_{\max}$
  are the number of temperatures, the minimum and the maximum
  temperatures used in the Parallel Tempering method; $N_\mathrm{Met}$
  is the number of Metropolis sweeps (at each temperature);
  $\mathrm{ps/spin}$ is the average CPU time per spin-flip in MUSI-MSC,
  using an Intel Xeon CPU E5-2680 processors; $N_\mathrm{samp}$ denotes
  the number of bad samples whose simulations had to be extended in
  order to thermalize and finally $N_\mathrm{Met,min}$,
  $N_\mathrm{Met,mean}$ and $N_\mathrm{Met,max}$ are the minimum, mean and
  maximum number of Metropolis sweeps per temperature needed to reach
  thermalization (bad samples). The set of temperatures used is
  clearly the same in the MUSI-MSC and MUSA-MSC parts of this
  Table. The number of Metropolis sweeps between two consecutive
  Parallel Tempering sweeps is always $N_\mathrm{MpPT} = 10$.  For the
  MUSI-MSC simulation of $L=24$ we parallelized, using
  \emph{Pthreads}, by distributing the $N_T=24$ system copies among 12
  CPU cores in the Intel Xeon CPU E5-2680.}
\end{table}

\section{On the selection of relevant parameters of the simulation}
\label{app:selection_parameters}

%In this appendix we explain how certain simulation parameters have been chosen
%and we justify that our selection do not affects the results obtained.

\begin{figure}
\centering
%\hspace{-1cm}
\includegraphics[width=0.8\columnwidth]{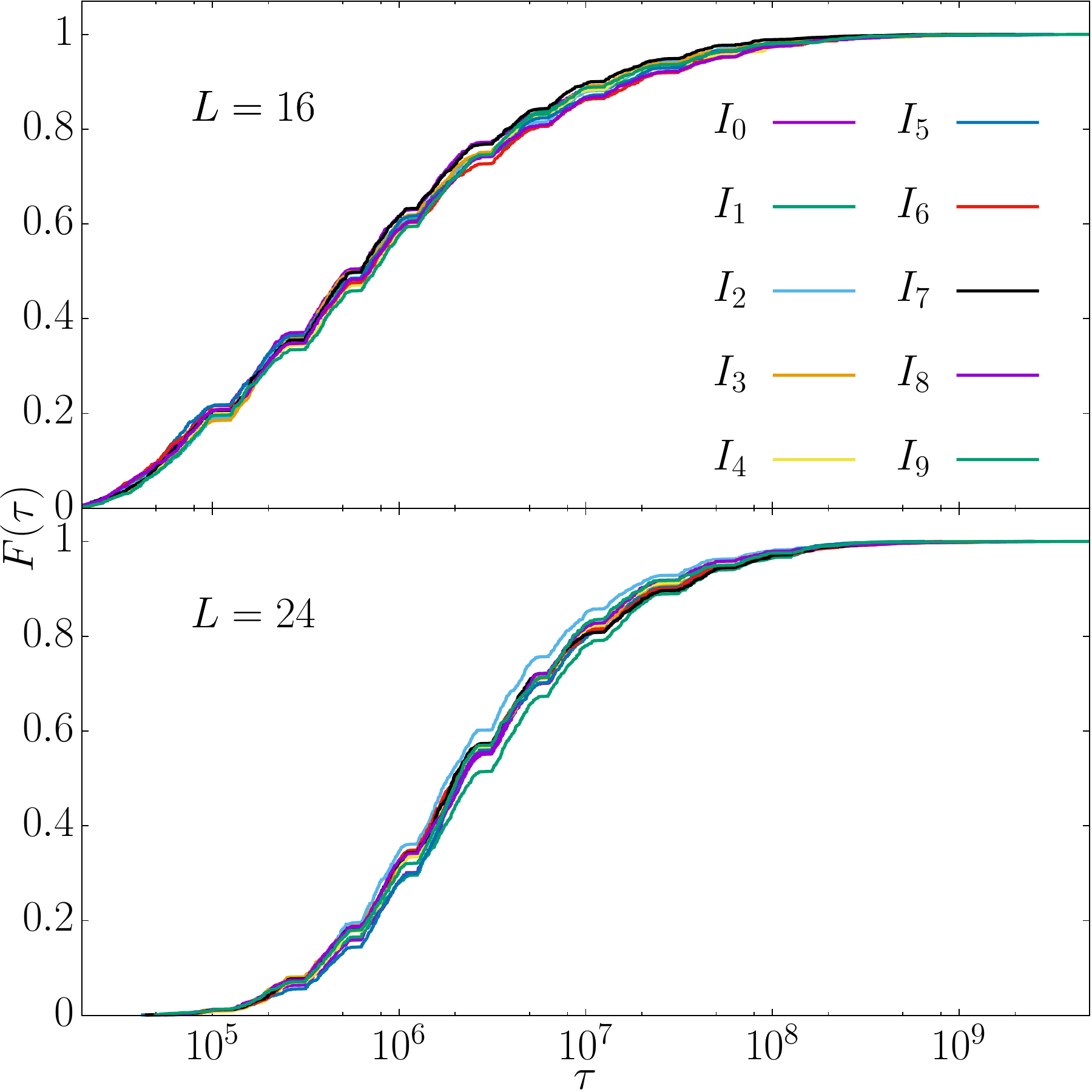}
\caption{\label{fig:selection_samples_dinamica} Empirical probability
  distribution function of $\tau$ represented for the $10$ inner
  samples separately.  $L=16$ case (top) and $L=24$ case
  (bottom). Averaging over the metastate (i.e. the outer samples) with
  fixed inner couplings reduces strongly the fluctuations between the
  inner samples.  }
\end{figure}

\begin{figure}
\centering
%\hspace{-1cm}
\includegraphics[width=0.8\columnwidth]{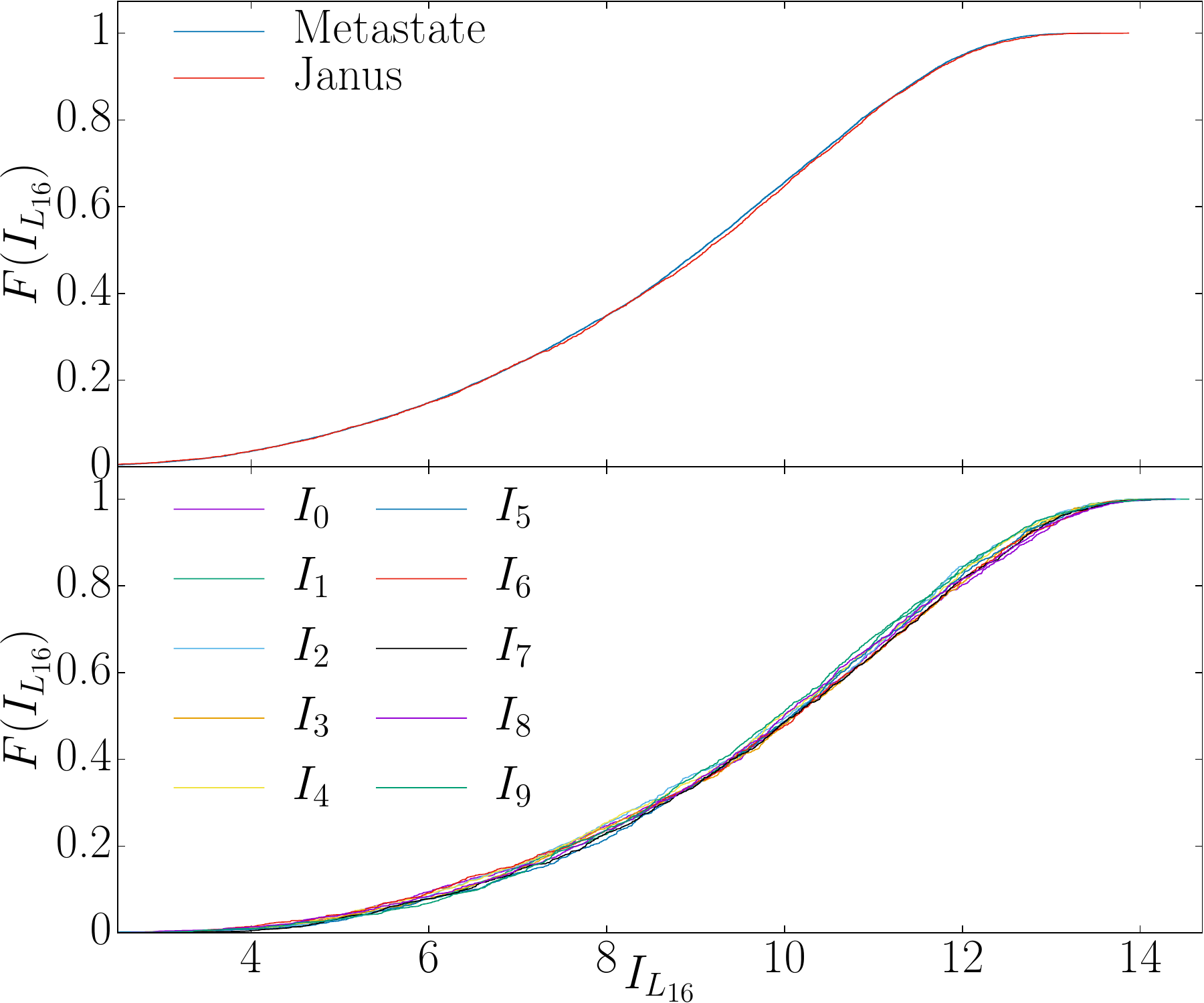}
\caption{\label{fig:selection_samples_estatica} Empirical probability
  distribution function of the integrated chaotic parameter. {\bf Top}
  We compare the distribution (labeled as "Metastate") obtained with
  our particular choice of samples with the distribution obtained from
  $4000$ fully independent samples (data from Janus). {\bf Bottom:}
  Distributions obtained for the $10$ inner samples plotted
  separately. Averaging over the metastate (over the outer couplings)
  reduces strongly the fluctuations between the inner samples.}
\end{figure}

The natural question is whether our particular choice of samples (see~\ref{app:simulation_parameters}) affects our results. 
One could imagine that the results obtained from configurations
sharing the same inner part could be strongly correlated, and that
with only $10$ inner parts, our statistics would be insufficient.  We
show in Fig \ref{fig:selection_samples_dinamica} that this is not the
case for the probability distribution of $\tau$: the probability
distributions of $\tau$ for the samples sharing the same $10$ inner
parts are plotted separately. they are nearly indistinguishable. The
average over the outer disorder (that we can call the metastate
average in analogy with Ref. \cite{billoire:17}) reduces dramatically
the fluctuations due to the inner disorder.  The same
conclusion holds for the chaos integral (see Fig.
\ref{fig:selection_samples_estatica})

\begin{figure}
\centering
%\hspace{-1cm}
\includegraphics[width=0.8\columnwidth]{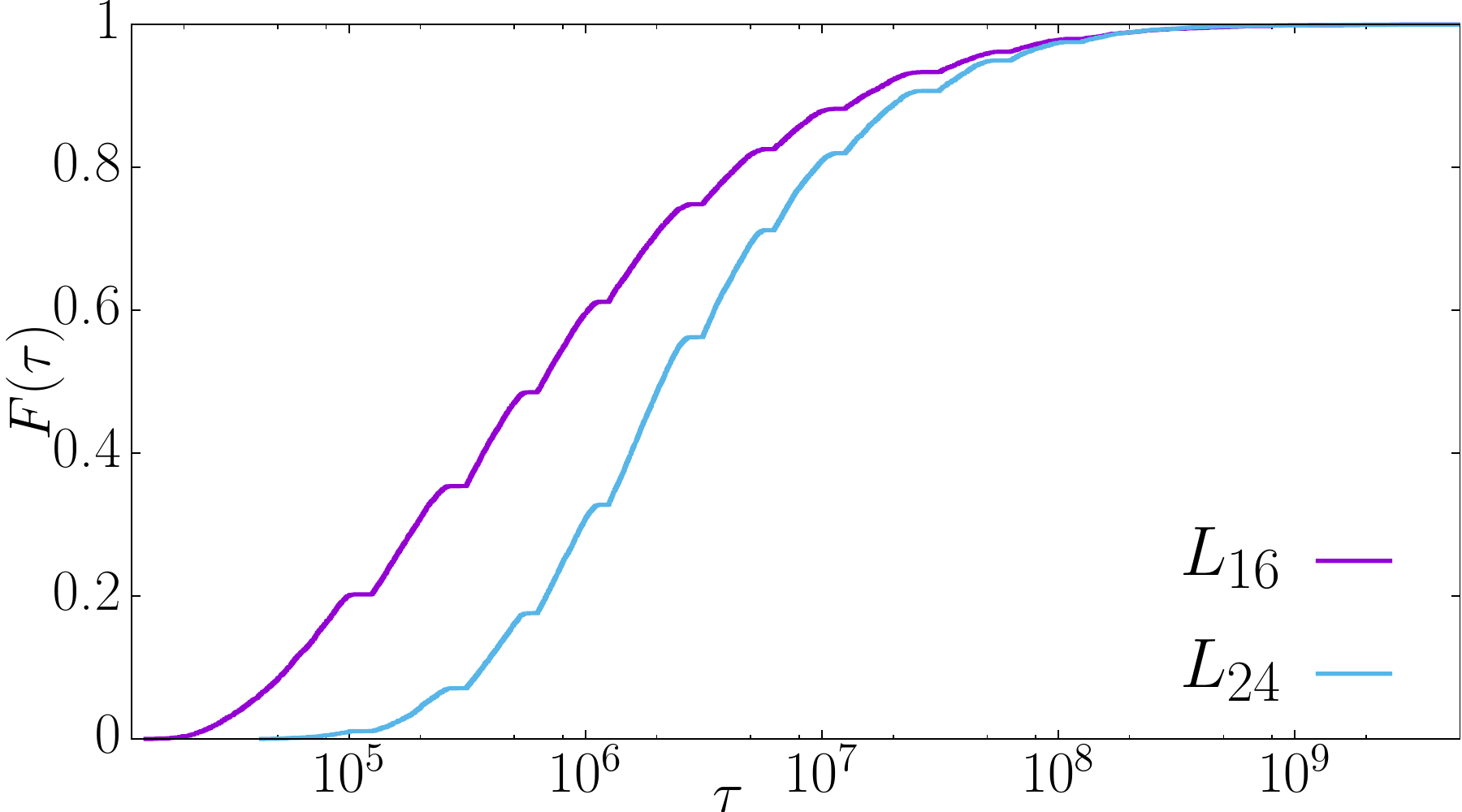}
\caption{\label{fig:comparison_dynamics_L24_L16} Empirical
  probability distribution function of $\tau$. Comparison of results
  for the simulations ($L=24$,$T_\mathrm{min}=0.698$) and
  ($L=16$,$T_\mathrm{min}=0.479$). Note that at the high-end of very
  difficult samples, these two simulations are similarly challenging.}
\end{figure}

On the other hand, the selection of the minimal temperature in the
Parallel Tempering could seem arbitrary, however the selection of
$T_\mathrm{min}^{L=16}$ and $T_\mathrm{min}^{L=24}$ have been made
carefully to assure that the most difficult samples had similar
$\tau$. This is shown in the figure
\ref{fig:comparison_dynamics_L24_L16}.

\section{The geometry of MUSI-MSC}\label{app:geometry_S_MSC}

The geometric construction explained in Ref.~\cite{fernandez:15} for
$L=256$ turns out to be satisfactory as well for $L=16$, but not for
$L=24$.  Hence, we shall first recall the geometry that we employ for
$L=16$. Afterwards, we explain the modifications that we introduced
for $L=24$. Note that multispin coding is not usually employed in
single-sample simulations because, in common schemes, one needs an
independent random number per bit. Fortunately, this problem can be
circumvented as explained in Ref.~\cite{fernandez:15}.

For $L=16$, the physical lattice of Cartesian coordinates $0\leq
x,y,z<L$ is mapped to a \emph{super-spin} lattice. Each super-spin is
coded in a 256-bits computer word (of course, the 256 bits correspond
to 256 physical spins which are updated in parallel). The crucial
requirement is that spins which are nearest-neighbors in the physical
lattice are  coded into nearest-neighbors super-spins.  In particular,
our super-spins are placed at the nodes of a cubic lattice with the
geometry of a parallelepiped of dimensions $L_x=L_y=L/8$, and
$L_z=L/4$. The relation between physical coordinates $(x,y,z)$ and the
coordinates in the super-spin lattice $(i_x,i_y,i_z)$ is
\begin{eqnarray}
x&=& b_x L_x + i_x\,,\ 0\leq i_x<L_x\,,\ 0\leq b_x < 8\,,\nonumber\\\label{eq:MSC-lattice}
y&=& b_y L_y + i_y\,,\ 0\leq i_y<L_y\,,\ 0\leq b_y < 8\,,\\
z&=& b_z L_z + i_z\,,\ 0\leq i_z<L_z\,,\ 0\leq b_z < 4\,.\nonumber
\end{eqnarray}
In this way, exactly 256 sites in the physical lattice are given the same
super-spin coordinates $(i_x,i_y,i_z)$. We differentiate between them by means of
the bit index:
\begin{equation}
i_b=64 b_z+8b_y+b_x\,,\ 0\leq i_b\leq 255\,.
\end{equation}
Since we have to simulate $N_T$ independent system copies in our Parallel
Tempering simulation, we simply carry out successively the simulation of
the $N_T$ systems. 

The alert reader will note that the above geometric construction is
very anisotropic (we start with a cube, but end-up with a
parallelepiped). Fortunately, this unsightly feature can be easily
fixed by noticing that the single-cubic lattice is bipartite. Indeed,
the lattice splits into the \emph{even} and \emph{odd} sub-lattices
according to the parity of $x+y+z$. The two sub-lattices contain
$L^3/2$ sites.  Furthermore, odd spins interact only with even spins
and vice versa.  It follows that the update ordering is irrelevant,
provide that our full-lattice sweep first updates all the (say) odd
sites and next all the even sites. Now, provided that $L_x$, $L_y$ and
$L_z$ are all \emph{even}, the parity of $x+y+z$ and $i_x+i_y+i_z$
coincide.  This implies that all the spins coded in a single super-spin
share the same parity, making irrelevant the super-spin lattice
asymmetry. For $L=16$ one finds that $L_x=L_y=2$ and $L_z=4$, the three of
them even numbers, and hence the above geometric construction works
smoothly. 

Unfortunately, for $L=24$ one has $L_x=L_y=3$ and $L_z=6$ which
implies that the super-spin lattice cannot be split into even and odd
sub-lattices. Our solution consisted in introducing \emph{logical}
super-spins of $512$ physical spins, that were later on coded into two
computer words of 256 bits each. The geometrical correspondence was
($L_x=L_y=L_z=L/8$)
\begin{eqnarray}
x&=& \tilde b_x L_x + j_x\,,\ 0\leq j_x<L_x\,,\ 0\leq \tilde b_x < 8\,,\nonumber\\\label{eq:MSC-lattice-L24}
y&=& \tilde b_y L_y + j_y\,,\ 0\leq j_y<L_y\,,\ 0\leq \tilde b_y < 8\,,\\
z&=& \tilde b_z L_z + j_z\,,\ 0\leq j_z<L_z\,,\ 0\leq \tilde b_z < 8\,.\nonumber
\end{eqnarray}
In this way, exactly 512 sites in the physical lattice are given the same
super-spin coordinates $(j_x,j_y,j_z)$. We differentiate between them by means of
the bit index:
\begin{equation}
j_b=64 \tilde b_z+8\tilde b_y+\tilde b_x\,,\ 0\leq i_b\leq 511\,.
\end{equation}
Now, the crucial observation is that (because $L_x=L_y=L_z=3$ for
$L=24$), the parity of $x+y+z$ coincides with that of $j_x+j_y+j_z$ if
(and only if) the parity of $\tilde b_x+\tilde b_y+\tilde b_z$ is
even. In other words, given super-spin coordinates $(j_x,j_y,j_z)$ the
512 spins coded in the super-spin split into 256 even spins and 256 odd
spins. Because same-parity spins are guaranteed to be mutually
non-interacting, we decided to code the 256 bits with the same parity
in the same computer word, with the corresponding bit index being the
integer part of $j_b/2$.

However, the acceleration obtained with the MUSI-MSC
 was not enough for some of the worse $L=24$ samples. Hence, we
decided to add an extra layer of parallelism by using {\em Pthreads\/}
to simulate a single sample in multicore processors. Given the
smallness of the super-spin lattice we found it preferable not to use
concurrent threads in the simulation of a single system copies (recall
that we have $N_T=24$ system copies in the Parallel Tempering
simulation of $L=24$). Rather, we distributed the $N_T$ system copies
along 12 CPU cores, achieving an average speed of $57$ picoseconds per spin-flip.

\section{Quantities not related to chaos}\label{app:quatities_not_chaos}

\begin{figure}
\centering
%\hspace{-1cm}
\includegraphics[width=0.8\columnwidth]{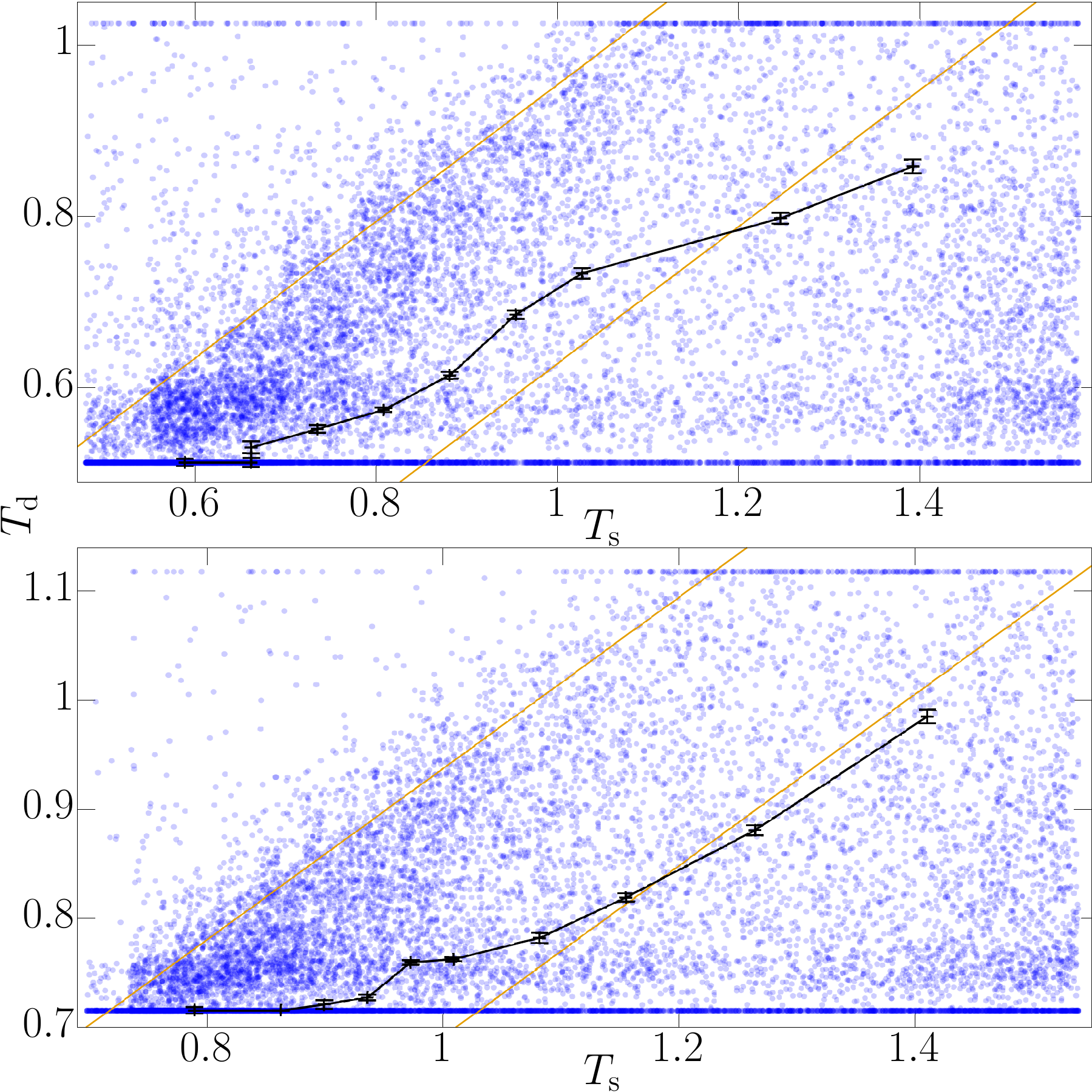}
\caption{\label{fig:Td_Ts} Scatter plot of $T_d$ versus $T_s$. We
  present the $L=16$-data (top) and the $L=24$-ones (bottom). Points
  are calculated with an special procedure. First, samples are
  classified on deciles according to $\log(\tau_{\mathrm{int}})$. The
  points coordinates were obtained by computing the median $T_d$ and
  the median $T_s$ within each decile (errors from bootstrap). The red
  parallel lines enclose the area of over-density that presents a
  higher correlation for later recount.}
\end{figure}

\begin{table}[h]
\centering
\begin{tabular*}{0.8\columnwidth}{@{\extracolsep{\fill}} ccccc}
\br
&$L$  & & $r$ & \\
\mr
&$16$ & & $0.348 \pm 0.008$ &\\
&$24$ & & $ 0.342 \pm 0.007$ & \\
\br
\end{tabular*}

\caption{\label{tab:coef_correla_TdTs} Correlation coefficients of the
  scatter plot of $T_d$ against $T_s$ for the simulated two lattice
  sizes.}
\end{table}

Some perfectly reasonable quantities turn out to have surprisingly
little relation to temperature chaos. To illustrate this effect, we
test whether or not the temperature obtained through the variational
method $T_\mathrm{d}=\lbrace T^* :
\tau_\mathrm{int}=\tau_\mathrm{int,var}\rbrace$ is correlated with the
static temperature of chaos $T_\mathrm{s}$ (see Fig.  \ref{fig:Td_Ts}).

In this case, Fig. \ref{fig:Td_Ts} shows an over-density, however the
points out of the principal density are too dispersed. For $L=16$
(top) the number of points within the lines are $8017$ ($62.63 \%$ of
the total) while for $L=24$ (bottom) the number of points within the
lines are $7539$ ($58.90 \%$ of the total). If we calculate the
correlation coefficients we obtain the Table
\ref{tab:coef_correla_TdTs}.

\section{Analyzing parallel tempering simulations}\label{app:Markov}

In the main text we have used theoretical tools to analyze the time
series produced by a Markov chain~\cite{sokal:97} in a setting that might
be unfamiliar in the context of Statistical Mechanics.  In
particular, in our parallel tempering simulations we have a number
$N_T$ of independent copies (or \emph{clones}) of the spin system that
we want to simulate. Each clone wanders along the \emph{temperature
  axis}, and our analysis focused \emph{solely} on these temperature
excursions. At first sight, the reader might be surprised by the fact
that this \emph{temperature wandering} may teach us something about
how far the spins are from thermal equilibrium at each
temperature. The purpose of this appendix is to briefly clarify the
relationship between both types of degrees of freedom, namely the
clone temperatures and the spins (see also Refs.  \cite{marinari:98b,
  marinari:98g, katzgraber:06b, fernandez:09b, janus:10}).

For sake of clarity, this appendix is organized in three paragraphs.
A Markov Chain Monte Carlo describes a random-walk 
process: in~\ref{sect:espacio_de_fases} we describe the phase space where
our random-walk takes place. We also discuss in~\ref{sect:espacio_de_fases} the stationary probability
distribution (i.e. the \emph{equilibrium} distribution) that our
random walk is targeted to reach. In~\ref{sect:Markov_chain-generator} we analyze some basic facts
about the dynamics of a Markov process (see for example
Ref.~\cite{sokal:97} for a more detailed discussion). Finally in~\ref{sect:example} we consider an example that will hopefully
clarify the matter further.

%%%%%%%%%%%%%%%%%%%%%%%%%%%%%%%%%%%%%%%%%%%%%%%%%%%%%%%%%%%%%%%%%%%%%%%%%%
\subsection{The phase space and the equilibrium distribution}\label{sect:espacio_de_fases}

We consider a cubic lattice of linear size $L$ with periodic boundary
conditions. We define a set of $N_T$ temperatures, with $T_1 < T_2
<\cdots < T_{N_T}$. Our random walk moves in a discrete, very large
phase space.
Each state point, denoted $X$, $Y$, $Z$\ldots
hereafter, is composed of two elements.
\begin{itemize}
\item The spins: for each lattice site $x$ we have $N_T$ binary
  variables $s_x^{(\alpha)}=\pm 1$. Here, $\alpha$ is the clone
  index, which takes values $\alpha=1,\ldots,N_T$.
\item The clone permutation $\pi$: $\pi$  is a permutation of
  $N_T$ symbols (there are $N_T!$ such permutations). The action of
  the permutation over the clone index $\alpha$, $\pi(\alpha)$, has a
  simple interpretation: it means that clone $\alpha$ is currently at
  temperature $T_{\pi(\alpha)}$.
\end{itemize}
In order to emphasize the composite nature of our state-point, we use
the notation $X=\big\{ \pi, \{ s_x^{(\alpha)}\}_{\alpha
  =1}^{N_T}\big\}$. The state point can take $N_T!\,2^{N_T L^3}$
values. The position of the random walk in phase space depends on
time: $X_t=\big\{ \pi_t, \{
s_x^{(\alpha)}\}_{t,\alpha=1}^{N_T}\big\}$. The random walk has by
construction the stationary distribution
\begin{equation}\label{eq:Prob_equilibrioA}
  P_{\mathrm{eq}}(X) =  \frac{1}{N_T!} \prod_{\alpha=1}^{N_T} \frac{\mathrm{exp} [-H(\{s_x^{(\alpha)}\})/T_{\pi{(\alpha)}}]} {Z_{T_\pi{(\alpha)}}}\,,
\end{equation}
where $H$ is the Edwards-Anderson Hamiltonian defined in Eq.~\eqref{eq:H} and $Z_{T_\alpha}$ is the partition function at temperature
$T_\alpha$. One can also write it as
\begin{equation}\label{eq:Prob_equilibrioB}
P_{\mathrm{eq}}(X) =  
\frac{1}{N_T!} \prod_{\alpha=1}^{N_T} \frac{\mathrm{exp} [-H(\{s_x^{\pi^{-1}(\alpha)}\})/T_\alpha]} {Z_{T_\alpha}}\,,
\end{equation}
where $\pi^{-1}$ is the inverse permutation of $\pi$
($\pi(\pi^{-1}(\alpha))=\alpha$ for any $\alpha$).
Let us now consider the
conditional probability conditioned to a given value of $\pi$.
Without loss of generality we select
$\pi=\mathbbm{1}$,  the identity permutation, such that
$\mathbbm{1}(\alpha)=\alpha$ for all $\alpha$:
\begin{equation}
P_{\mathrm{eq}}(X|\pi=\mathbbm{1})=\frac{\mathrm{e}^{-H(\{s_x^{(1)}\})/T_{1}}} {Z_{T_1}}
\frac{\mathrm{e}^{-H(\{s_x^{(2)}\})/T_{2}}} {Z_{T_2}}\ldots
\frac{\mathrm{e}^{-H(\{s_x^{(N_T)}\})/T_{N_T}}} {Z_{T_{N_T}}}\,.
\end{equation}
This conditional probability is a product of distributions (i.e. the
spins for clones $\alpha\neq\beta$ are statistically independent,
provided that $\pi$ is kept fixed), and the equilibrium probability
distribution for the spins $\{s_x^{(\alpha)}\}$ is the Boltzmann
distribution for temperature $T_\alpha$.

Two marginal probabilities extracted from $P_{\mathrm{eq}}(X)$ 
are of interest:
\begin{itemize}
\item Tracing-out the spin degrees-of-freedom in
  Eq.~\eqref{eq:Prob_equilibrioA} one sees that the
  equilibrium probability for the clones permutation is uniform:
\begin{equation}
P_{\mathrm{eq},\mathrm{marginal}}(\pi)=\frac{1}{N_T!}\,. 
\end{equation}
Specializing to clone $\alpha$, we find
$P_\mathrm{eq}(\pi(\alpha)=\beta)=1/N_T$ for any $\beta$.
Checking that this has been achieved with good accuracy
for all clones is one of the important tests of thermalization.
\item The equilibrium probability for the spins of the clone currently
  at temperature $T_\beta$, namely $\alpha=\pi^{-1}(\beta)$, is
\begin{equation}
  P_{\mathrm{eq},\mathrm{marginal}}(\{s_x^{(\alpha)}\}|\pi(\alpha)=\beta) =
  \frac{\mathrm{exp} [-H(\{s_x^{(\alpha)}\})/T_{\beta}]}{Z_{T_\beta}}\,.
\end{equation}
In other words, when the random-walk equilibrates, Boltzmann
equilibrium is reached at all $N_T$ temperatures: the spin
configuration of the clone currently at temperature $T_\beta$ is a
typical configuration of the Boltzmann distribution at such temperature.
\end{itemize}

%%%%%%%%%%%%%%%%%%%%%%%%%%%%%%%%%%%%%%%%%%%%%%%%%%%%%%%%%%%%%%%%%%%%%%%%%
\subsection{The random walk and its correlation functions}
\label{sect:Markov_chain-generator}

We consider a stationary Markov process~\cite{sokal:97}.  When going
from time $t$ to time $t+1$ the system is updated $X_t\rightarrow
X_{t+1}$ with a time-independent rule, that only uses as input the
current state $X_t$. Previous states ($X_{t-1}$, $X_{t-2}$, $...$)
have no influence on the decision of where to move at time $t+1$.

In our case the Markov dynamics is generated by a square matrix
$G_{X,Y}$ of dimension $N_T!\,2^{N_T L^3}$ that meets two basic
conditions, namely $G_{X,Y}\geq 0$ and $\sum_X G_{X,Y}=1$.  In fact,
$G_{X,Y}$ is a conditional probability: it is the probability for
having $X_{t+1}=X$ when one knows that
$X_t=Y$.\footnote{Ref.~\cite{sokal:97} employs a reversed convention,
  where our $G_{X,Y}$ is named $T_{Y,X}$. As a consequence,
  Ref.~\cite{sokal:97} reverses the ordering of vector and matrices in
  matrix products, see e.g. Eq.~\eqref{eq:master-eq}.} It follows that the
probability for having $X_{t=k}=X$, namely $P_{t=k}(X)$, obeys the
master equation
\begin{equation}\label{eq:master-eq}
P_{t=k}(X)=\sum_Y [G^k]_{X,Y} P_{t=0}(Y)\,,
\end{equation}
where $G^k$ is the $k$-th power of the generating matrix $G$.  Matrix
$G$ is carefully crafted to fulfill the balance
condition\footnote{Specifically, our $G$ is factorized as
  $G=G_{\mathrm{Temperature\ Swap}}
  [G_{\mathrm{Metropolis}}]^{10}$. During the Metropolis part of the
  dynamics the spins of clone $\alpha$ evolve with a standard
  Metropolis dynamics at temperature $T_{\pi(\alpha)}$ (each factor
  $G_{\mathrm{Metropolis}}$ corresponds to a full-lattice sweep). The
  permutation $\pi$ is changed by matrix
  $G_{\mathrm{Temperature\ Swap}}$. We try to exchange sequentially
  $\pi^{-1}(\alpha)$ with $\pi^{-1}(\alpha+1)$, for $\alpha=1,2,\ldots N_T-1$
  (in this way, the clone at the lowest temperature has a theoretical
  chance to reach the highest temperature in a single Parallel
  Tempering iteration). Each temperature swap attempt is
  accepted or rejected according to a Metropolis test, see
  e.g.~Ref.~\cite{marinari:98g}.}
\begin{equation}\label{eq:balance}
\sum_Y G_{X,Y} P_{\mathrm{eq}}(Y)=P_{\mathrm{eq}}(X)\,.  
\end{equation}
The balance condition states that the equilibrium
distribution~\eqref{eq:Prob_equilibrioA} is a right-eigenvector of
matrix $G$, with eigenvalue 1. When combined with the master equation,
the balance condition tells us that the equilibrium distribution is a
stationary distribution for our random walk. 

Let us consider the spectral decomposition of the initial distribution
on the $N_T!2^{N_TL^3}$ right-eigenvectors of matrix $G$, $Gu_n=\lambda_n u_n$ 
(ordered in such a way that $1>|\lambda_1|>|\lambda_2|>...$):
\begin{equation}
P_{t=0}=P_{\mathrm{eq}}+\sum_{n} c_n u_n\,.
\end{equation}
The master equation implies that
\begin{equation}\label{eq:equilibration}
P_{t=k}=P_{\mathrm{eq}}+\sum_{n} c_n \lambda_n^k u_n\,.
\end{equation}
Hence, $P_{t=k}$ converges exponentially to $P_{\mathrm{eq}}$ and the
corresponding exponential auto-correlation time is
$\tau_\mathrm{exp}=-1/\mathrm{log} |\lambda_1|$. 

However, the spectral analysis of the equilibrium correlation
functions (see Sect.~\ref{sect:tau-def}) is carried in terms of the left
eigenvectors of matrix $G$, $\tilde u_n G=\lambda_k \tilde
u_k$. Fortunately, for any matrix left-eigenvalues coincide with
right-eigenvalues (instead, left and right
eigenvectors typically differ). In fact, these are the eigenvalues appearing in
Eq.~\eqref{eq:eigen-1}, that we repeat here for the reader's
convenience
\begin{equation}\label{eq-appendix:eigen-1}
\hat C_f(t)=\sum_n A_{n,f} \lambda_n^{|t|}\,,\quad \sum_n A_{n,f}=1\,.
\end{equation}
In particular, the constant vector $\tilde u_0$ 
[$\tilde u_0(X)=1$ for all states $X$] is a left eigenvector with eigenvalue 1.
The generic observable $f$ considered in Eq.~\eqref{eq-appendix:eigen-1}
can be decomposed as
\begin{equation}
f(X)=E(f) \tilde u_0(X) \ +\  \sum_n B_{n,f} \tilde u_n(X)\,,
\end{equation}
where $E(f)$ is the \emph{equilibrium} expectation value.  The
coefficients $A_{n,f}$ in Eq.~\eqref{eq-appendix:eigen-1} are
$A_{n,f}= \tilde B_{n,f}/(\sum_{n'} \tilde B_{n',f})$, where $\tilde
B_{n,f}=B_{n,f}E\big(\tilde u_n(X)[f(X)-E(f)]\big)$.

The crucial message from this analysis is that the characteristic time
scales $\tau_n$.\footnote{Remember that $\lambda_n=\mathrm{e}^{-1/\tau_n}$}
that one identifies by studying
the correlation functions, as we did in the main text, are
\emph{exactly} the timescales that govern the approach to equilibrium,
see Eq.~\eqref{eq:equilibration}. These characteristic times $\tau_n$
can be obtained from \emph{any} convenient observable $f$. Whether $f$
is a spin observable, or something related to the clone permutation is
immaterial. The only thing that really matters is that $A_{n=1,f}$
should be as large as possible.

%%%%%%%%%%%%%%%%%%%%%%%%%%%%%%%%%%%%%%%%%%%%%%%%%%%%%%%%%%%%%%%%%%%%%%%
\subsection{An example}\label{sect:example}

Just to show how deeply the spin and the temperature dynamics are
intertwined, we consider here in details an example. We shall consider
a typical $L=24$ sample instance (neither extremely easy, nor
extremely hard: it roughly corresponds to percentile 90 of hardness,
see Fig.~\ref{fig:all_L_prob_tau}).

We consider the standard parallel tempering simulation protocol from
the main text: $N_T=24$, $T_\mathrm{min}=0.698$. For this particular
sample one needs to run the simulation for $2\times 10^9$ Metropolis
sweeps (for each clone) in order to meet our thermalization
criteria. We also consider a truncated simulation where we only keep
the lowest four temperatures: $N_T=4$, $T_\mathrm{min}=0.698$,
$T_2=0.735$, $T_3=0.771$ and $T_4=0.808$ (all four deep in the
spin-glass phase, since $T_\mathrm{c}=1.102(3)$~\cite{janus:13}). The
truncated simulation is also run for $2\times 10^9$ Metropolis sweeps
per clone.

Our expectation is that the standard simulation will equilibrate,
while the truncated simulation will not. The rationale for this
expectation is simple: in the standard simulation, each clone spends
some $2\times 10^9/24\approx 8\times 10^7$ Monte Carlo steps at the highest
temperature. Yet, the exponential auto-correlation time for the
Metropolis dynamics at $T=1.6$ is about $10^4$ lattice
sweeps~\cite{ogielski:85}. Hence, the time spent by each clone at the
highest temperature is long enough to effectively de-correlate the
system. Instead, the highest temperature in the truncated simulation
$T_{\mathrm{max,truncated}}=0.808$ lies well below $T_\mathrm{c}$. At
such a low temperature, the Metropolis dynamics is too inefficient to
decorrelate the system in only $2\times 10^9/4=5\times 10^8$
Metropolis sweeps.

Besides the temperature dynamics already considered in the main text,
we shall also study here the dynamics of spin observables. Using
the fact that we have already equilibrated this sample, we have
selected randomly four equilibrium spin configurations at our
lowest temperature $T_\mathrm{min}=0.698$, $\{\tau_{x,a}\}$
$a=1,2,3,4$. Then, for each clone, we compute the time-dependent
overlap
\begin{equation}\label{eq:overlap-mask}
q_{a,\alpha}(t)=\frac{1}{L^3} \sum_x \tau_{x,a} s_x^{(\alpha)}(t)\,.
\end{equation}
We always compute the overlap with a given clone $\alpha$, irrespective
its time-dependent temperature $T_{\pi_t(\alpha)}$.

We compute the overlaps $q_{a,\alpha}(t)$ from a set of ten new
standard simulations ($N_T=24$), with a random start, where we measure
the overlaps very often (every $5\times 10^4$ Metropolis sweeps). We
also compute the overlaps $q_{a,\alpha}(t)$ from our new truncated
simulation with $N_T=4$ (the truncated simulation had a random start,
as well).  Recall that, as we said above, the spin masks
$\{\tau_{x,a}\}$ are taken from the previous sets of simulations that
were discussed in the main text.

\begin{figure}[t]
\centering
\includegraphics[width=0.8\columnwidth]{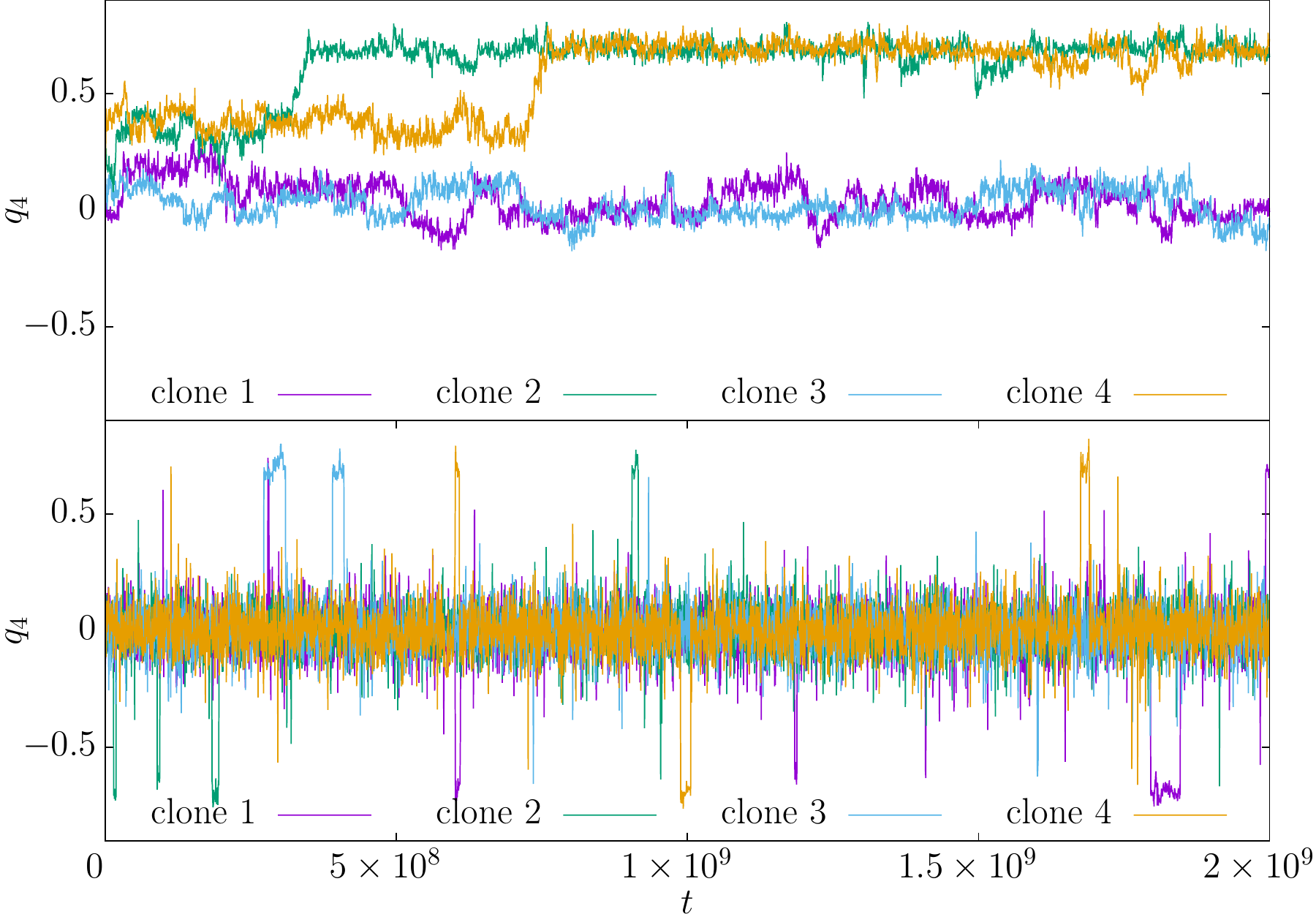}
\caption{\label{fig:app_historia_q} {\bf Top:} Monte Carlo history for
  the overlap $q_4(t)$, see Eq.~\eqref{eq:overlap-mask}, as computed
  for each of the four clones in the truncated simulation (see
  text). Note that our simulation time is much too short to expose the
  symmetry $q_4 \leftrightarrow -q_4$. As a consequence, we know for
  sure that thermal equilibrium has not been reached for the truncated
  simulation.  {\bf Bottom:} As in top panel, for the first four
  clones in one of our standard simulations with $N_T=24$ temperatures
  (there were 10, completely independent, standard simulations). For
  each clone, the overlap $q_4(t)$ changes sign many times along the
  simulation (as it is to be expected for a well equilibrated
  simulation). Note that, with small probability, each clone reaches a
  state where $|q_4|\sim 0.8$. These events, which are not observed
  for the other three overlaps $q_a$ $a=1,2,3$, make particularly
  interesting to study the dynamics of $q_4$.  }
\end{figure}

The global spin flip symmetry of the Edwards-Anderson Hamiltonian
implies that the equilibrium distribution for $q_{a,\alpha}$ is
symmetric under $q_{a,\alpha}\leftrightarrow -q_{a,\alpha}$. It is
important to check this symmetry, since it is believed that the
largest dynamical barriers are related to global
spin-flips~\cite{billoire:01}.\footnote{The alert reader will point
  out that the eigenvectors of the dynamical matrix $G$ can be
  classified according to their parity with respect to global
  spin-flip symmetry. However, because spin-flip symmetry is
  spontaneously broken in the low temperature-phase, spin-flip
  transitions are exponentially (in some power of $L$) suppressed in
  local Monte Carlo at fixed temperature. The only efficient mechanism
  for producing a global spin-reversal is having the clone travel to
  the high-temperature end of the Parallel Tempering temperature
  grid.}

The Monte Carlo history of the time-dependent overlap with $\tau_4$,
that we call $q_4$ in Fig.~\ref{fig:app_historia_q}, shows very
clearly that the truncated simulation is not able to reach thermal
equilibrium within the time span of our simulations. The (not shown)
Monte Carlo history for the other overlaps, $q_{a=1,2,3}$ are
qualitatively similar. Instead, the standard simulation displays the
expected symmetry under $q_{4,\alpha}\leftrightarrow
-q_{4,\alpha}$. The Monte Carlo histories (in the standard simulation)
for $q_{a,\alpha}$ with $a=1,2,3$ (not shown) are symmetric as
well. Only $q_4$ uncovers a state that arises with small probability,
characterized by $|q_4|\sim 0.8$. This feature suggests that $q_4$ is
the most interesting overlap to look at.

\begin{figure}
\centering
\includegraphics[width=0.8\columnwidth]{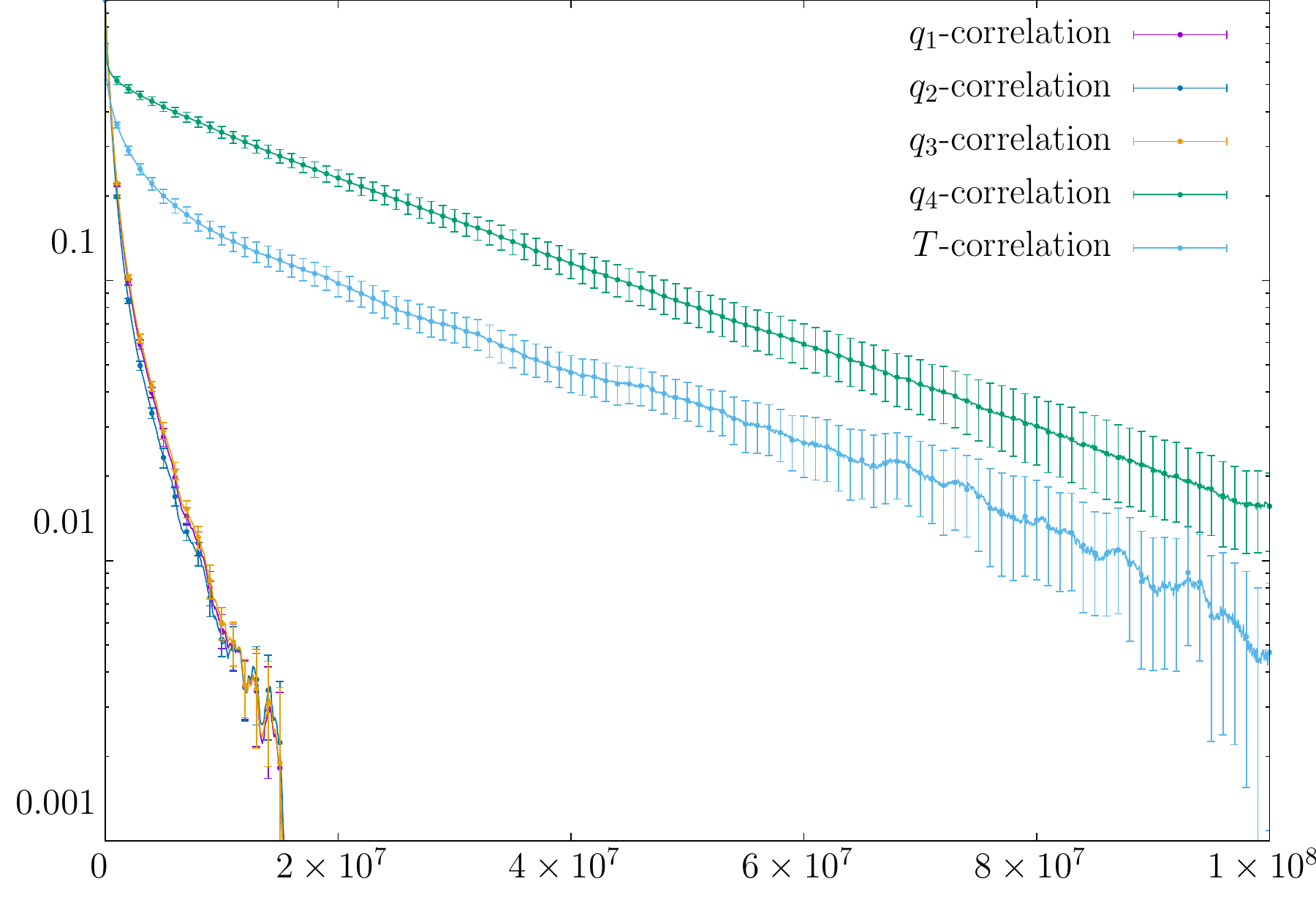}
\caption{\label{fig:app_correlaciones} Equilibrium time dependent
  correlation functions, as computed from the standard simulation with
  $N_T=24$.  We consider five observables, one related to temperature
  (computed from the piece-wise linear function with $T^*=T_3$, see
  Table~\ref{tab:table_functions} and Sect.~\ref{sect:tau-var}), and
  the overlaps $q_a$ with $a=1,2,3,4$ defined in
  Eq.~\eqref{eq:overlap-mask}. The fact that the $T$ and $q_4$
  correlations become parallel in this semi-logarithmic scale
  indicates that we are safely computing the exponential
  auto-correlation time (which is independent of the
  observable). Instead, the $q_{a=1,2,3}$ correlations do not
  become parallel to the other curves, at least not within the range
  we can measure, which probably indicates that the amplitudes
  $A_{n=1,q_{a=1,2,3}}$, see Eq.~\eqref{eq-appendix:eigen-1}, are much
  smaller for these observables.}
\end{figure}

In order to make the above impressions quantitative, we show in
Fig.~\ref{fig:app_correlaciones} some equilibrium correlation
functions, which can be computed, of course, only for the standard
simulation. As it could be expected from
Sect.~\ref{sect:Markov_chain-generator}, the very same exponential
auto-correlation time is computed from the temperature random walk, or
from the $q_4$ correlation (specifically, and measuring time in
Metropolis sweeps, we find $10^{-7}\tau_{\mathrm{exp}}=3.0(4)$ from
$q_4$, while we find the fully compatible value
$10^{-7}\tau_{\mathrm{exp}}=3.1(6)$ from the $T$ random-walk). One could
conclude from Fig.~\ref{fig:app_correlaciones} that the computation of
$\tau_{\mathrm{exp}}$ is simpler by considering $q_4$ than by studying
the temperature random walk. This is a misleading conclusion, though:
we had to equilibrate the system, in the first place, in order to find
the spin mask $\{\tau_{x,a=4}\}$ that defines the overlap $q_4$.
Furthermore, the other spin masks $\{\tau_{x,a=1,2,3}\}$, turned out
not to be particularly useful in the computation of the exponential
auto-correlation time. It is in no way guaranteed that one can
identify an interesting overlap by picking randomly a small number of
equilibrated configurations.

\begin{figure}
\centering
\includegraphics[width=0.8\columnwidth]{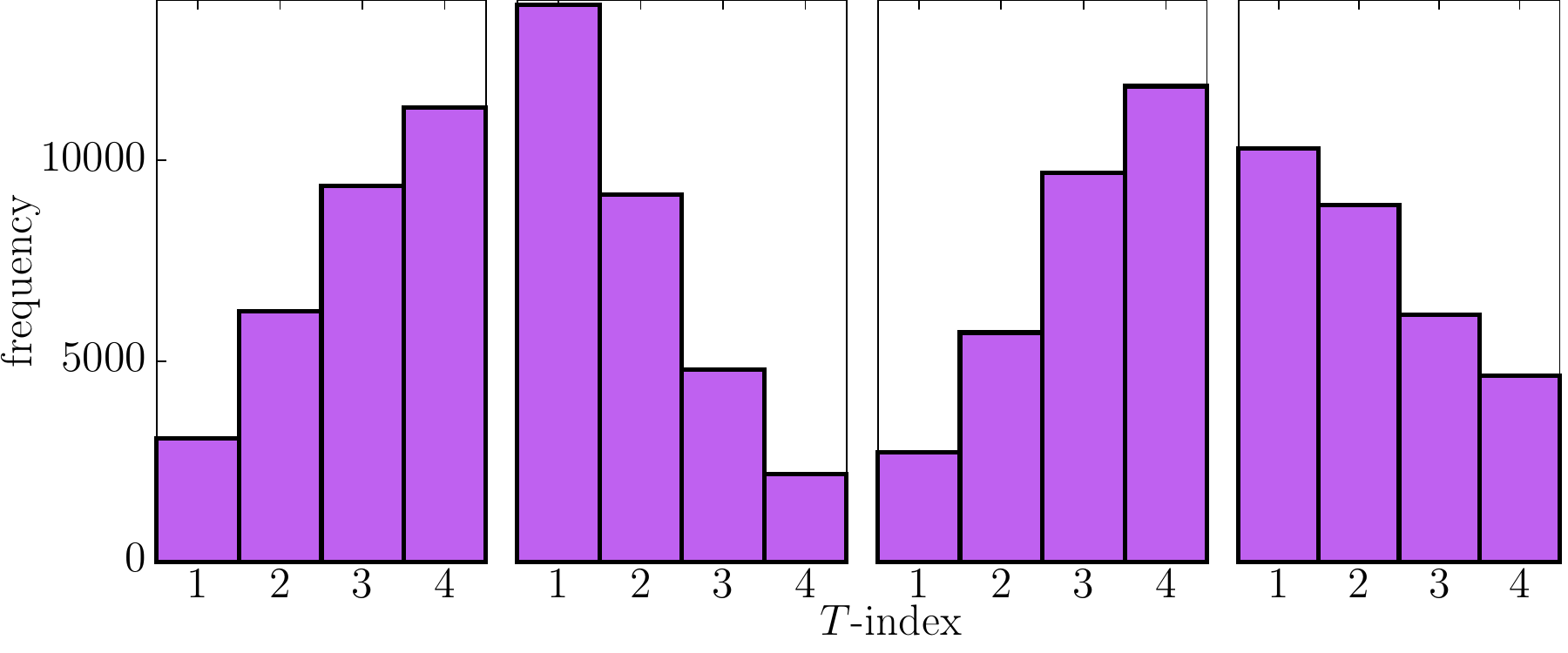}
\caption{\label{fig:app_histograma} For each of the four clones in the
  truncated simulation, we indicate the histogram of temperature
  (i.e. the number of times that $\pi_t(\alpha)=1$, or
  $\pi_t(\alpha)=2$, etc.). The temperature state was sampled every
  $5\times 10^4$ Metropolis sweeps (per clone).  Had the simulation
  equilibrated, we would have expected the occupation histograms to be
  uniform.}
\end{figure}

Finally, one could consider a different
question. Fig.~\ref{fig:app_historia_q} shows beyond any question that
the truncated simulation does not reach equilibrium. However, there
are only 4 clones in that run and one could believe that it should be
not that difficult to equilibrate the clone permutation. The question
is investigated in Fig.~\ref{fig:app_histograma} by means of an
occupation histogram (it is not possible to compute equilibrium
correlation functions for a simulation that does equilibrate). The
answer to our query is an unqualified no: the fact that the spins are
out from equilibrium makes it also impossible to equilibrate the clone
permutations.

\section*{References}

%\bibliographystyle{iopart-num} 
%\bibliography{/homenfs/rg/biblio.bib}

\providecommand{\newblock}{}

\end{document}